\documentclass[%
reprint,
superscriptaddress, 
amsmath,amssymb,
aps,
prl,
prstab,
floatfix,
longbibliography
]{revtex4-2}

\usepackage[caption=false]{subfig}
\usepackage{multirow}
\usepackage{graphicx}
\usepackage{dcolumn}
\usepackage{bm}
\usepackage{threeparttable}
\usepackage{color}
\usepackage[dvipsnames]{xcolor}
\usepackage{chemformula}
\usepackage{xspace}
\usepackage{orcidlink}
\usepackage{amsmath}
\usepackage[normalem]{ulem}

\definecolor{myblue}{RGB}{55, 126, 184}

\newcommand{\RMF}{$\mathrm{RbMnF}_{3}$}
\newcommand{\KNF}{$\mathrm{KNiF}_{3}$}

\definecolor{AF}{RGB}{0, 0, 0}
\definecolor{AMK}{RGB}{0, 0, 0}
\definecolor{AVK}{RGB}{0, 0, 0}
\definecolor{RD}{RGB}{0, 0, 0}
\definecolor{OLD}{RGB}{0, 0, 0}
\definecolor{JM}{RGB}{1, 0, 0}
\definecolor{DA}{RGB}{220, 0, 0}
\definecolor{TTG}{RGB}{255, 0, 0}
\definecolor{Old}{RGB}{192,192,192}

\bibpunct{[}{]}{;}{n}{}{}

\hypersetup{colorlinks=true, linkcolor=blue, citecolor=blue, urlcolor=blue, pdftitle = {{C}oherent {THz} {S}pin {D}ynamics in {A}ntiferromagnets {B}eyond the {M}acrospin {A}pproximation}, pdfauthor = {T.T.~Gareev}}

\begin{document}

\title{{C}oherent {THz} {S}pin {D}ynamics in {A}ntiferromagnets \\{B}eyond the {A}pproximation of the {N}\'eel vector}

\author{F.~Formisano\,\orcidlink{0000-0001-8789-4518}}

     \affiliation{Institute for Molecules and Materials, Radboud University, 6525 AJ Nijmegen, The Netherlands}
     
\author{T.~T.~Gareev\,\orcidlink{0000-0002-1678-2444}}
     \affiliation{Institute for Molecules and Materials, Radboud University, 6525 AJ Nijmegen, The Netherlands}

\author{D.~I.~Khusyainov\,\orcidlink{0000-0003-1332-4146}}
     \affiliation{Institute for Molecules and Materials, Radboud University, 6525 AJ Nijmegen, The Netherlands}

\author{A.~E.~Fedianin\,\orcidlink{0000-0002-5093-6830}}
     \affiliation{Ioffe Institute, Russian Academy of Sciences, 194021 St.\,Petersburg, Russia}
     
\author{R.~M.~Dubrovin\,\orcidlink{0000-0002-7235-7805}}
     \affiliation{Ioffe Institute, Russian Academy of Sciences, 194021 St.\,Petersburg, Russia}

\author{P.~P.~Syrnikov}
     \affiliation{Ioffe Institute, Russian Academy of Sciences, 194021 St.\,Petersburg, Russia}

\author{D.~Afanasiev\,\orcidlink{0000-0001-6726-3479}}
     \affiliation{Institute for Molecules and Materials, Radboud University, 6525 AJ Nijmegen, The Netherlands}

\author{R.~V.~Pisarev\,\orcidlink{0000-0002-2008-9335}}
\affiliation{Ioffe Institute, Russian Academy of Sciences, 194021 St.\,Petersburg, Russia}

\author{A.~M.~Kalashnikova\,\orcidlink{0000-0001-5635-6186}}
\affiliation{Ioffe Institute, Russian Academy of Sciences, 194021 St.\,Petersburg, Russia}

\author{J.~H.~Mentink\,\orcidlink{0000-0002-4082-1984}}     \affiliation{Institute for Molecules and Materials, Radboud University, 6525 AJ Nijmegen, The Netherlands}

\author{A.~V.~Kimel\,\orcidlink{0000-0002-0709-042X}}
\email{aleksei.kimel@ru.nl}
\affiliation{Institute for Molecules and Materials, Radboud University, 6525 AJ Nijmegen, The Netherlands}

\date{\today}

\begin{abstract}
Controlled generation of coherent spin waves with highest possible frequencies and the shortest possible wavelengths is a cornerstone of spintronics and magnonics.
Here, using the Heisenberg antiferromagnet \RMF{}, we demonstrate that laser-induced THz spin dynamics corresponding to pairs of mutually coherent counter propagating spin waves with the wavevectors up to the edge of the Brillouin zone cannot be understood in terms of magnetization and antiferromagnetic (N\'eel) vectors, conventionally used to describe spin waves.
Instead, we propose to model such spin dynamics using the spin correlation function.
We derive a quantum-mechanical equation of motion for the latter and emphasize that, unlike the magnetization and antiferromagnetic vectors the spin correlations in antiferromagnets do not exhibit inertia.

\end{abstract}

\maketitle

Exploring efficient routes for excitation of coherent spin waves with the shortest possible wavelength and the highest achievable frequency is one of the major challenges of today's spintronics, magnonics, and magnetic data storage~\cite{pirro2021advances,che2020efficient,csaba2017perspectives,mentink2015ultrafast,sandweg2011spin}.
Among magnetic materials, antiferromagnets have the highest frequencies of spin dynamics in the THz range, which is a significant advantage for becoming a crucial ingredient of ultra-high speed spintronic devices~\cite{barman2021roadmap,fukami2020antiferromagnetic,nemec2018antiferromagnetic,jungwirth2018multiple,baltz2018antiferromagnetic,jungwirth2016antiferromagnetic}. 
It has been demonstrated that ultrashort laser pulses can be employed for the generation of coherent spin waves in practically all classes of magnetically ordered materials \cite{kalashnikova2007impulsive,gridnev2008phenomenological,satoh2015writing,tzschaschel2017ultrafast}. 
The wavelength of the optically excited spin wave is, in principle, defined by the size of the illuminated volume, and can be further decreased by introducing inhomogeneities~\cite{satoh2012spinwaves,afanasiev2021spinwaves}.
However, the spin waves thus generated will still have frequencies and wavevectors close to the center of the Brillouin zone.
An appealing alternative approach for the generation of spin waves with larger k-vectors is based on concomitant excitation of the two counter-propagating waves~\cite{kotyuzhanskii1981parametric,sandweg2011spin}.
In antiferromagnets, such mutually coupled pairs of spin waves excited over the whole Brillouin zone form a so-called two-magnon mode, which couples to light via Raman scattering \cite{fleury1968scattering}, and can be triggered by ultrashort laser pulses \cite{zhao2004magnon,bossini2016macrospin}.

Until now, the approximation, which neglects individual \textcolor{AMK}{strongly coupled quantum mechanical} spins and treats a magnet as a continuous medium, has been fundamental to understanding spin dynamics, including that at ultrafast timescales \cite{kimel2020fundamentals}.
Using such an approximation, the simplest two-sublattice antiferromagnet can be modeled as two antiferromagnetically coupled ferromagnets with oppositely directed magnetizations $|\mathbf{M}_{1}|$ = $|\mathbf{M}_{2}|$ = $M_{0}$, the dimensionless antiferromagnetic vector $\mathbf{L}$ = ($\mathbf{M}_{1}$ - $\mathbf{M}_{2}$)/$2M_{0}$ and the net magnetization $\mathbf{M}$ = ($\mathbf{M}_{1}$ + $\mathbf{M}_{2}$)/$2M_{0}$. 
The propagation of an antiferromagnetic spin wave with the wavevector $\mathbf{k}$ and the angular frequency $\Omega_\mathbf{k}$ in this approximation is described as spatio-temporal variations of the orientations and lengths of \textbf{L} and \textbf{M}.
It is clear that upon decreasing the wavelength and approaching the edge of the Brillouin zone, the approximation based on such macroscopic vectors should eventually fail \textcolor{AMK}{since it is substantiated only if the characteristic length scale of spin wave is much larger than the interspin distance.}
Despite this fact, the breakdown has not been examined for experimentally observed laser-induced spin dynamics of antiferromagnets. 

Here, using the example of the Heisenberg antiferromagnet \RMF, we demonstrate that the spin dynamics corresponding to mutually coherent spin waves with large opposite wavevectors cannot be understood if the light-spin interaction is modeled in conventional terms of macroscopic magnetization $\mathbf{M}$ and antiferromagnetic vector $\mathbf{L}$. 
Instead, we propose to model such spin dynamics using spin correlations. 
We derive the equation of motion for the spin correlation function and demonstrate that, unlike $\mathbf{M}$ and $\mathbf{L}$, the spin correlations in antiferromagnets do not exhibit inertia. 
It means that there is an optimal pulse duration for excitation of the spin correlation dynamics, while the macrospin dynamics in the antiferromagnet can be excited impulsively even by an infinitesimally short pulse. 

\begin{figure}
 \centering
 \includegraphics[width=1\columnwidth]{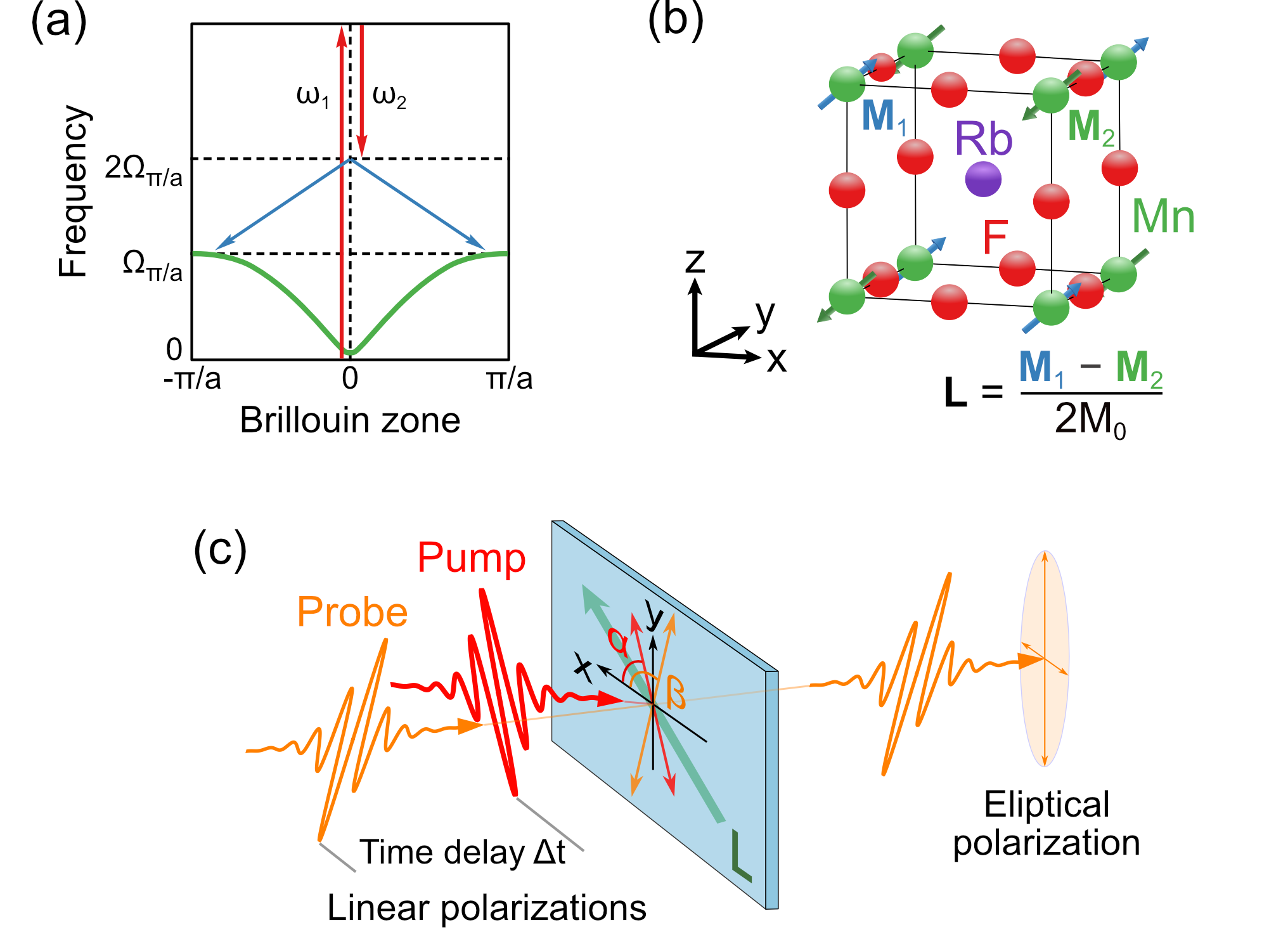}
 \caption{\label{fig:experimental_scheme}
 (a)~Coupling of two photons with frequencies differing by $2\Omega_\mathbf{k}=\omega_{1}-\omega_{2}$ with a pair of the coherent spin waves with frequencies $\Omega_\mathbf{k}$ and opposite wavevectors $\mathbf{k}$ throughout the Brillouin zone.
 The dominant contribution to the process from the spin waves at the edge of the Brillouin zone is shown.
 (b)~Crystallographic and magnetic structure of \ch{RbMnF3}.
 The arrows indicate antiferromagnetically ordered spins aligned along the $\langle111\rangle$ direction.
 (c)~Experimental geometry.
 Linearly polarized optical pump pulse excites spin dynamics which is detected via transinet ellipticity of the optical probe pulse time-delayed by $\Delta t$. 
 The projection of the antiferromagnetic vector $\mathbf{L}$ on the $xy$ sample plane is shown.
 }
 \end{figure}

The excitation of the two-magnon mode relies on the perturbation of exchange coupling, and the symmetry of the latter is intrinsically linked to the crystal structure, while for the models in terms of \textbf{M} and \textbf{L} an orientation of magnetic moments is important.
Therefore, it was crucial to choose a material in which the orientation of the magnetic moments and the resulting antiferromagnetic vector $\mathbf{L}$ are distinctly different from the main crystallographic axes.
Here we select the cubic fluoroperovskite insulator \RMF{} (point group $m\overline{3}m$), which is very close to the isotropic 3D Heisenberg antiferromagnet below the N{\'e}el temperature $T_{N}=83$\,K~\cite{teaney1962discovery,freiser1963field}. 
The spins ($S=5/2$) of the $\mathrm{Mn}^{2+}$ ions form two equivalent magnetic sublattices coupled antiferromagnetically. 
Very weak magnetic anisotropy aligns the antiferromagnetic vector $\mathbf{L}$ along one of the $\langle111\rangle$ directions~\cite{lopez2014magnetic}, as shown in Fig.~\ref{fig:experimental_scheme}(b).
The two-magnon mode in \ch{RbMnF3} was observed in spontaneous Raman scattering experiments [Fig.~\ref{fig:experimental_scheme}(a)]; has the $E_{g}$ symmetry and frequency 4\,THz at low temperature~\cite{fleury1968evidence,elliott1969effects,fleury1970temperature,barocchi1978determination}.
 
To excite coherent spin dynamics corresponding to the two-magnon mode, we perform a femtosecond two-color magneto-optical pump-probe experiment in the transmission geometry, as shown in Fig.~\ref{fig:experimental_scheme}(c).
The laser pulses with duration 45~fs and central photon energy 1.55~eV were emitted by the Ti:Sapphire regenerative amplifier at a repetition rate of 1\,kHz and were split into pump and probe ($\hbar\omega_{pr}=1.55$~eV) beams with an energy per pulse ratio of 10:1.
The pump pulses were tuned by an optical parametric amplifier (OPA) to a central 
photon energy $\hbar\omega_{p} = 1.03$~eV and a fluence of 1.2\,mJ/cm$^{2}$ and then were focused normally onto the sample surface into a spot with a diameter of 500\,$\mu$m. 
The pump and probe pulses were linearly polarized at angles $\alpha$ and $\beta$ with respect to the $x\|[100]$ axis, respectively. 
The angles were controlled by half-wave plates.
The 620\,$\mu$m-thick slab sample with the optical quality of the surfaces was cut perpendicular to the cubic $z\|[001]$ axis from a single crystal of \RMF{} grown using the Czochralski method. 
The projection of the antiferromagnetic vector $\mathbf{L}$ on the sample $xy$ plane makes an angle of 45$^\circ$ with the $x-$axis, as shown in Fig.~\ref{fig:experimental_scheme}(c).
The sample was placed in the continuous-flow liquid helium cryostat. 
The spin dynamics triggered by an intense pump pulse induces in the sample a certain type of optical anisotropy \cite{satoh2015writing}.
The latter is detected using a balance detection scheme to measure the induced ellipticity of the polarization of the probe pulse as a function of the time delay $\Delta{t}$ between pump and probe pulses.
Note that in this detection scheme, only modulation of the medium parameters which is nearly homogeneous within the probed area is detected.

\begin{figure}
 \centering
 \includegraphics[width=1\columnwidth]{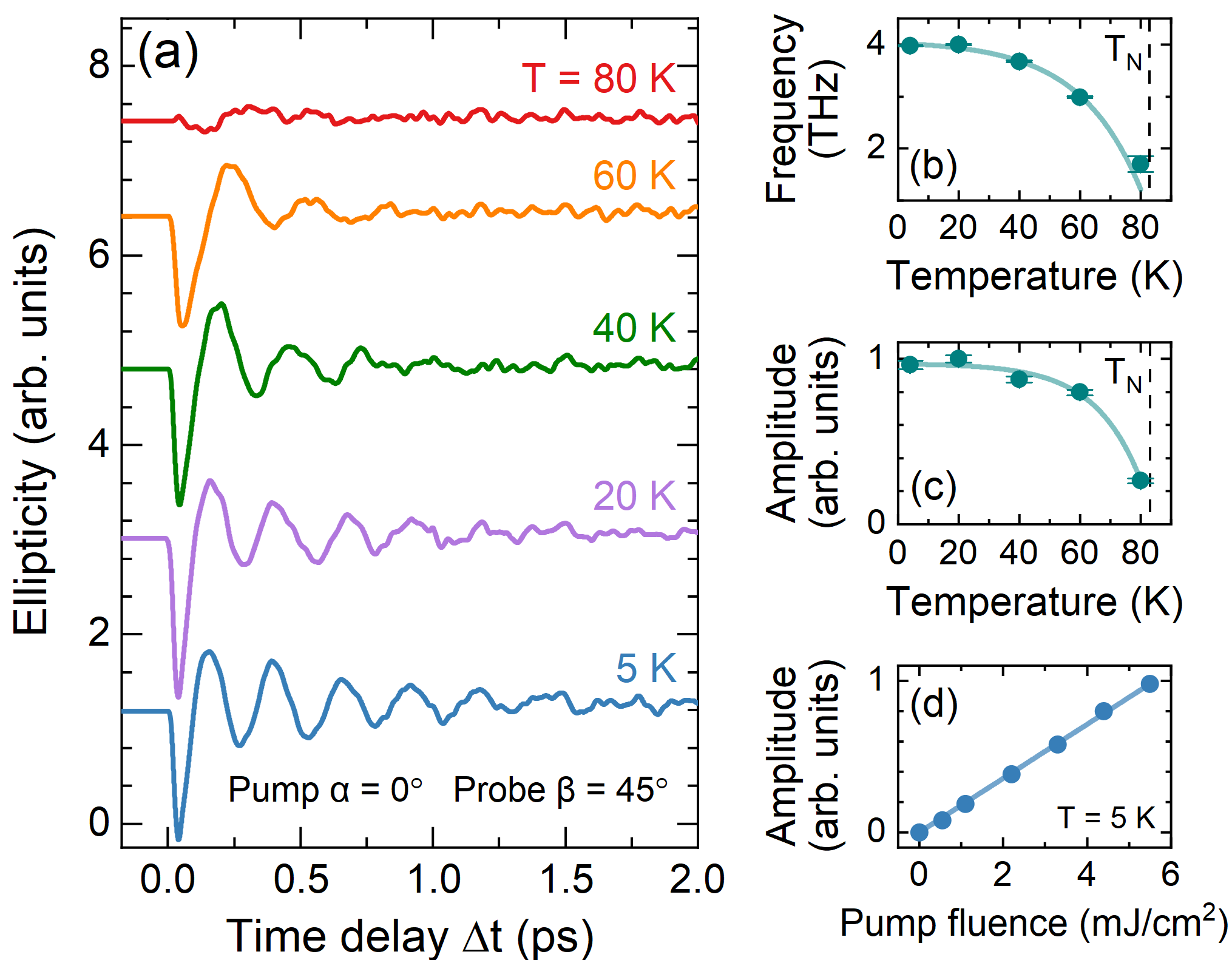} 
 \caption{\label{fig:ellipticity_temperature}
 (a)~Transient probe ellipticity at different temperatures below $T_{N}=83$\,K.
 Pump and probe pulses are linearly polarized at angles $\alpha=0^{\circ}$ and $\beta=45^{\circ}$, respectively.
 Temperature dependences of the (b)~frequency and (c)~amplitude of the oscillations seen in (a).
 (d)~Dependence of the amplitude of the oscillations on the pump pulse fluence at $T=5$~K.
 The solid lines in (b-d) are guides for the eye.}
 \end{figure}

Figure~\ref{fig:ellipticity_temperature}(a) shows a temporal evolution of the pump-induced ellipticity in the antiferromagnetic phase of \RMF.
The pump and probe pulses polarizations were $\alpha=0^\circ$ and $\beta=45^\circ$. 
One clearly distinguishes oscillations damped within a few ps.
The Fourier spectra of the time trace for $T=5$\,K reveals a clear resonance at 4\,THz. 
The frequency and amplitude of these oscillations decrease upon heating up to $T_{N}$ [Figs.~\ref{fig:ellipticity_temperature}(b,c)].
The frequency and its temperature behavior are in fair agreement with those of the two-magnon mode obtained in spontaneous Raman scattering experiments~\cite{fleury1968evidence,elliott1969effects,fleury1970temperature,barocchi1978determination}.
The detected two-magnon mode corresponds to the simultaneous excitation of pairs of mutually coherent spin waves with opposite wavevectors throughout the Brillouin zone dominated by the waves with the shortest wavelengths and the highest frequencies $\Omega_{\pi/a}$, as shown in Fig.~\ref{fig:experimental_scheme}(a)~\cite{fleury1968scattering,fleury1968evidence,pisarev1973light,lockwood1992one,meloche2007two}.
Thus, we can confidently conclude that the experimentally detected oscillations correspond to the spin dynamics originating from the pairs of spin waves with opposite wavevectors at the edges of the Brillouin zone. 

\begin{figure}
 \centering
 \includegraphics[width=1\columnwidth]{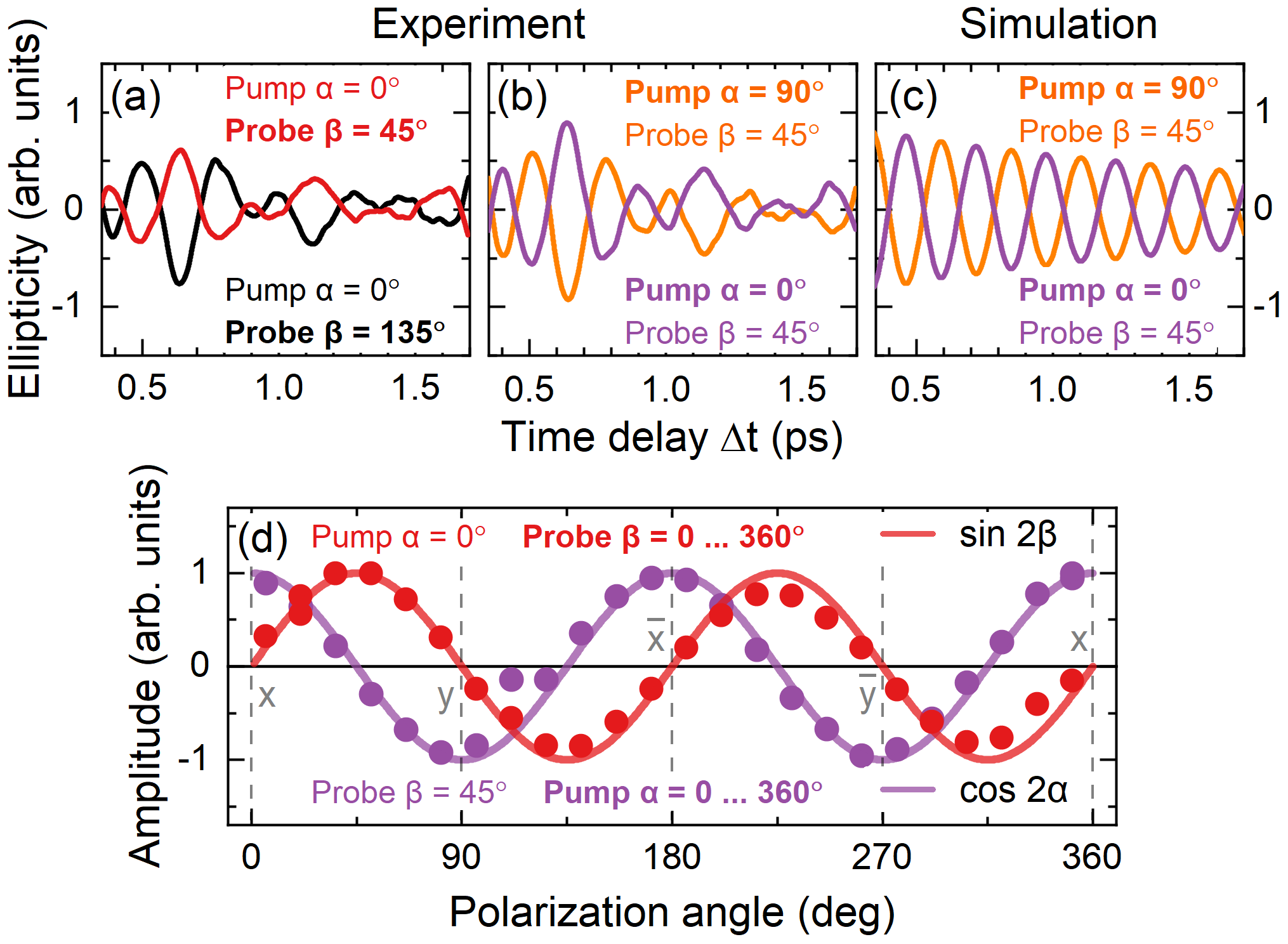}
 \caption{\label{fig:ellipticity_polarization}
 (a,b) Experimental and (c) calculated transient probe ellipticity as excited and detected at (a) fixed pump polarization $\alpha=0$ and the probe polarizations $\beta=45,135^\circ$, (b,c) fixed probe polarization $\beta=45^\circ$ and the pump polarizations $\alpha=0,90^\circ$.
 (d)~The signed Fourier amplitude of 4~THz oscillations as a function of pump polarization angle $\alpha$ at the probe polarization angle $\beta=45^{\circ}$ (purple symbols), and probe polarization angle $\beta$ at the pump polarization angle $\alpha=0^{\circ}$ (red symbols).
 The sign change corresponds to the waveform phase shift by $\pi$.
 The purple and red lines show the fit of the experimental data by $\cos2\alpha$ and $\sin2\beta$ functions, respectively.
 The experimental data are obtained at $T=5$\,K.
 }
 \end{figure}

To comprehend the observed spin dynamics, we performed measurements of the pump-induced ellipticity at various angles of the pump $\alpha$ and probe $\beta$ polarizations at $T=5$~K [Fig.~\ref{fig:ellipticity_polarization}].
The most pronounced oscillations of the probe ellipticity are observed when the pump polarization is along the crystallographic $x$ and $y$ axes ($\alpha = 0,~90^{\circ}$) and the probe polarization is at an angle $\beta = 45^{\circ},~135^{\circ}$. 
Moreover, it is clearly seen that the phase of the oscillations is shifted by $\pi$ when either pump or probe polarization is rotated by $90^\circ$ [Figs.~\ref{fig:ellipticity_polarization}(a,b)].
To gain further insight, we performed two series of experiments for fixed pump ($\alpha=0^{\circ}$) and rotated probe ($\beta=0^{\circ}\ldots360^{\circ}$) polarization and for fixed probe ($\beta=45^{\circ}$) and rotated pump ($\alpha=0^{\circ}\ldots360^{\circ}$) polarizations.
The variations of the signed Fourier amplitudes of the obtained waveforms are fairly well described by $\cos{2\alpha}$ and $\sin{2\beta}$ [Fig.~\ref{fig:ellipticity_polarization}(d)].

In contrast to earlier reported results of ultrafast spin dynamics \cite{kimel2020fundamentals}, the observed polarization dependencies cannot be explained in terms of the macroscopic magnetization $\mathbf{M}$ and antiferromagnetic vector $\mathbf{L}$.
In particular, it was argued that excitation of pairs of mutually coherent spin waves with opposite wavevectors in isostructural antiferromagnetic fluoroperovskite \KNF~\cite{bossini2019laser} modulate the length of the antiferromagnetic vector $\mathbf{L}$ at the two-magnon frequency homogeneously across the excitation area. 
The latter is detected in experiments as oscillations of magnetic linear birefringence at the same frequency.
In \KNF{} $\mathbf{L}$ is along one of the main crystallographic axes, e.g., the $x$ axis, while in the case of \RMF, the projection of $\mathbf{L}$ on the $xy$ sample plane is rotated over $45^{\circ}$ from the $x$ and $y$ axes. 
Nevertheless, the pump polarization dependences are exactly the same in both antiferromagnets and possess maxima when the pump is polarized along one of the main crystallographic axes.
Moreover, as the maximum signal in \RMF{} is observed when the probe is initially polarized along or perpendicular to the projection of $\mathbf{L}$ on the sample plane, the corresponding optical anisotropy does not originate from the oscillations of the length of $\mathbf{L}$ and cannot be understood in terms of magnetic linear birefringence.
The probing geometry could agree with the transverse dynamics of $\mathbf{L}$ detected via such effect, which, however, yields the paradoxical conclusion that $\mathbf{L}$ homogeneously oscillates at the frequency twice as high as the maximal spin wave frequency (Suppl. Mater.~\cite{supp_mat}).

To describe the observed polarization dependences, we have to go beyond the conventional approach in terms of dynamics of \textbf{L}.
We assume the antiferromagnetic Heisenberg exchange interaction with \textcolor{AMK}{Hamiltonian} $\hat{H}_0 = J \sum_{\boldsymbol{\delta}} \hat{C}(\boldsymbol{\delta})$, where $J$ is the exchange coupling constant between the nearest neighbor magnetic ions, $\boldsymbol{\delta}$ is the vector between them. 
For a cubic crystal, $\hat{C}(\boldsymbol{\delta})=2\sum_{i}\hat{\mathbf{S}}(\mathbf{r}_i) \hat{\mathbf{S}}(\mathbf{r}_i+\boldsymbol{\delta})$, \textcolor{AMK}{where $\mathbf{r}_i$ is a vector to site $i$.}
\textcolor{AMK}{The expectation value of $\hat{C}(\boldsymbol{\delta})$ gives the equal time spin correlation $\langle \hat S(\mathbf{r}_i,t)\hat S(\mathbf{r}_i+\boldsymbol{\delta},t)\rangle$}
along $\boldsymbol{\delta}$, \textcolor{AMK}{where $\langle\ldots\rangle$ denotes averaging}
~\cite{fedianin2022selection}. 
\textcolor{AMK}{For a cubic Mott insulator,} the electric field $\mathbf{E}=\left(E^x,E^y,E^z\right)$ of light propagating along the $z-$axis and polarized in the $xy-$plane perturbs the Hamiltonian according to the phenomenological expression~\cite{mentink2015ultrafast}
\begin{align}
    \Delta\hat{H}&= A(\omega_{p}) \left[ \left(E^x \delta^x \right)^2  \hat{C}(\delta^x) + \left(E^y \delta^y \right)^2 \hat{C}(\delta^y)\right],\label{eq:DeltaH}
\end{align}
where $A(\omega) \delta^\nu \delta^\nu \hat{C}(\delta^\nu)$ is proportional to the dielectric permittivity at frequency $\omega$ with spectral dependence $A(\omega)$ defined by exchange coupling, optical bandgap, and crystal structure.
\textcolor{AMK}{We note that Eq.~\eqref{eq:DeltaH} agrees with the expression for the perturbation of the Hamiltonian in the case of two-magnon Raman scattering in an antiferromagnet~\cite{elliott1969effects}.}

To derive the equation for the dynamics of spin correlations, we consider small transverse spin-deflections with respect to the collinear antiferromagnetic ground state. It is convenient to represent these spin deflections in reciprocal space, since spin correlations for different bonds are not independent: $[\hat{C}_i(\delta^\nu),\hat{C}_{j}(\delta^\mu)]\neq\delta_{ij}\delta_{\nu\mu}$, with $\delta_{ij}$ being a Kronecker delta. 
Indeed, spin correlations can share the same spin operator, for example, $\hat{\mathbf{S}}(\mathbf{r}_i+\boldsymbol{\delta}^\nu)=\hat{\mathbf{S}}(\mathbf{r}_j+\boldsymbol{\delta}^\mu)$. 
This means that perturbation of exchange bonds along the $x$ axis also triggers dynamics of spin correlations along the $y$ and $z$ axes, etc. 
Although the spin correlation itself is a scalar, not all components of the spin correlation feature the same dynamics. Analogously to the antiferromagnetic vector $\mathbf{L}$, we therefore write the different components of the spin correlation as a vector $\hat{\mathbf{C}}_\mathbf{k}=\left(\hat{C}_\mathbf{k}^{(1)},\hat{C}_\mathbf{k}^{(2)},\hat{C}_\mathbf{k}^{(3)}\right)$, where
\begin{align}
    \hat{C}^{(1)}_\mathbf{k}&=\frac{1}{2S}\left(\hat{S}^X_\mathbf{k} \hat{S}^X_\mathbf{-k}+\hat{S}^Y_\mathbf{k} \hat{S}^Y_\mathbf{-k}\right),\nonumber\\
    \hat{C}^{(2)}_\mathbf{k}&=\frac{1}{2S}\left(\hat{S}^X_\mathbf{k} \hat{S}^Y_\mathbf{-k}-\hat{S}^Y_\mathbf{k} \hat{S}^X_\mathbf{-k}\right),\\
    \hat{C}^{(3)}_\mathbf{k}&=\frac{1}{4S} \left((\hat{S}^X_\mathbf{k})^2+(\hat{S}^Y_\mathbf{k})^2 + (\hat{S}^X_\mathbf{-k})^2 + (\hat{S}^Y_\mathbf{-k})^2\right),\nonumber
\end{align}
with $X,Y$ being mutually orthogonal components perpendicular to the spin quantization axis $Z$.
Here we kept only the leading order, i.e., quadratic spin deflections~\cite{supp_mat} and static terms $\propto -S^2$ are omitted, which is justified by the linear dependence of the observed effect on the pump fluence [Fig.~\ref{fig:ellipticity_temperature}(d)].
Further, although temperature is not included in the theory, the results described below do not rely on explicit evaluation of the equations in the $T=0$ ground state. 
Note that in equilibrium, $\langle \hat{C}^{(2)}_\mathbf{k}\rangle =0$. However, it can be nonzero during dynamics. In terms of the vector components, the leading order deflections to the full correlation are given by
\begin{align}
\hat{C}_{\mathbf{k}}(\delta^\nu)=4S\left(\hat{C}^{(3)}_{\mathbf{k}}+\cos (k^\nu \delta^\nu)\hat{C}^{(1)}_{\mathbf{k}}\right),
\label{eq:Motion}
\end{align}
which shows that the correlation function features \textcolor{AMK}{a term $\propto\hat{C}^{(1)}$} not present in the dynamics of $\Delta L^Z\propto \sum_\mathbf{k}\langle\hat{C}^{(3)}_{\mathbf{k}}\rangle$~~\cite{bossini2019laser}.
\textcolor{AMK}{The dependence of the correlation on $\boldsymbol{\delta}$ stems entirely from the $\propto\hat{C}^{(1)}$ term, that was ignored before.}
From the commutations relations for the spin operators, it follows that 
$\hat{\mathbf{C}}_\mathbf{k}$ satisfies the cross product in hyperbolic space, $\hat{\mathbf{C}}_\mathbf{k}\times \hat{\mathbf{C}}_\mathbf{k}=i\hbar\hat{\mathbf{C}}_\mathbf{k}$, featuring a minus sign for the terms in the $\hat{C}^{(3)}_\mathbf{k}$ component as compared to the ordinary cross product~\cite{fabiani2022parametrically}.
Hence, from Heisenberg equations of motion, the dynamics of the correlations in reciprocal space is determined by
\begin{align}
\label{eq:C}
&\hbar\frac{\partial \hat{\mathbf{C}}_{\mathbf{k}}}{\partial t} - \hat{\mathbf{C}}_{\mathbf{k}} \times \left(-\frac{\partial\hat{H}_{\mathbf{k}}}{\partial \hat{\mathbf{C}}_{\mathbf{k}}}\right)=0,
\end{align}
where
\begin{align}\label{eq:Hamil}
&\hat{H}_{\mathbf{k}}= \left(J +A(\omega_{p}) \left(E^x \delta^x \right)^2 \right)\hat{C}_{\mathbf{k}}(\delta^x)\nonumber\\
&\quad\quad + \left(J +A(\omega_{p}) \left(E^y \delta^y \right)^2 \right)\hat{C}_{\mathbf{k}}(\delta^y)
+ J \hat{C}_{\mathbf{k}}(\delta^z).
\end{align}
Solving this equation of motion for the impulsive perturbation $\Delta\hat{H}$ gives precession of $\hat{\mathbf{C}}_{\mathbf{k}}$ with dominating frequency $\approx 2 z J S/\hbar$ ($z=6$ is the number of nearest neighbors), which corresponds to the frequency of the two-magnon mode [see Fig.~\ref{fig:ellipticity_polarization}(c) and Eqs.~(19,21) in Suppl. Mater.~\cite{supp_mat}]. 
Thus, we derived the equation of motion for spin correlations which describes the dynamics of this parameter, in agreement with our and previously reported experiments. 
We note, that, unlike \textbf{M} and \textbf{L} in antiferromagnets \cite{kimel2015inertia}, the dynamics of spin correlation is described by the differential equation of the first order and hence the spin correlations do not show inertia.

From Eq.~\eqref{eq:C}, we see that the polarization dependence of the pump-induced spin correlations is dictated by the light-induced perturbations of the effective fields $-{\partial\hat{H}_{\mathbf{k}}}/{\partial \hat{\mathbf{C}}_{\mathbf{k}}}$. 
In Suppl. Mater.~\cite{supp_mat} we demonstrate that for the experimental geometry considered one then obtains a polarization dependence $\propto\cos{2\alpha}$, in agreement with what is observed experimentally [Fig.~\ref{fig:ellipticity_polarization}(d)]. 
Phenomenologically, this polarization dependence can be understood as follows. 
From Eq.~\eqref{eq:DeltaH} one sees that an inequality $E^\nu\neq 0$ gives $\partial\Delta\hat{H}/\partial\hat{C}(\delta^\nu)\neq 0$. 
It means that pumping the antiferromagnet by a laser pulse polarized along one of the main crystallographic axes brings the system out of the thermodynamic equilibrium, and finding the new equilibrium will result in a change of the spin correlations. 
Such changes in the spin correlation functions contribute to the dielectric permittivity as $\Delta\varepsilon^{\mu\nu}=\partial^2\langle\Delta \hat{H}\rangle/\partial E^\nu\partial E^\mu$, which may lead to a pump-induced anisotropy $\varepsilon^{xx}-\varepsilon^{yy}\neq0$ in the initially optically isotropic cubic system. 
For the $m\overline{3}m$ point group, we obtain the pump-induced change of the dielectric permittivity $\varepsilon^{xx}-\varepsilon^{yy}=2gE^2\cos{2\alpha}$, where $g=\xi^{xxxx}-\xi^{yyxx}$, and $\boldsymbol{\xi}$ is the phenomenological polar tensor~\cite{birss1964symmetry}. 
In the range of low absorption, such pump-induced change of the dielectric permittivity results in the ellipticity of the probe pulse which obeys $\sin2\beta$ dependence on the incoming probe polarization, observed in the experiment [Fig.~\ref{fig:ellipticity_polarization}(d)].
Hence, the model based on spin correlation reproduces the experimental observation in terms of symmetry of excitation and detection. 

Furthermore, the model correctly describes the linear dependence of the effect on the pump pulse fluence [Fig.~\ref{fig:ellipticity_temperature}(d)], since the perturbation of the effective field in Eq.~\eqref{eq:C} is \textcolor{AMK}{linear in the pump pulse fluence $\propto (E^\nu)^2$}.
This contrasts with the result one obtains considering dynamics of \textbf{L}.
Indeed, the spin wave frequencies $\Omega_\mathbf{k}$ are much lower than those of the pump photons $\omega$, and the spin wave excitation involves a nonlinear process of down conversion $\omega_1-\omega_2=2\Omega_k$ [see Fig.~\ref{fig:experimental_scheme}(a)].
Hence, in the lowest order, a light-induced torque acting on $\mathbf{M}$ or $\mathbf{L}$ and the amplitude of the excited spin wave are \textcolor{AMK}{linear with respect to the pump fluence}, as observed for the optically-driven spin waves with low \textbf{k}~\cite{kalashnikova2007impulsive,kimel2015inertia}.
As the spin wave wavevectors $k\approx\pi/a$ are much larger than the wavevector of light $q\approx0$, they can be detected only in pairs such that $\mathbf{k}-\mathbf{k}=\mathbf{q}\approx 0$. 
Therefore, the signal detected by the probe pulse can only originate from a product of the spin-wave amplitudes. 
Hence, the detected signal must be quadratic with respect to the pump fluence (see Suppl. Mater.~\cite{supp_mat}).

In our model [Eqs.~(\ref{eq:DeltaH},\ref{eq:C})], when the electric field of the pump pulse is along one of the main crystallographic axes, it off-resonantly excites virtual charge transfer transitions between $\mathrm{Mn}^{2+}$ and $\mathrm{F}^{1-}$ ions and, thus, effectively changes the exchange interaction between $\mathrm{Mn}^{2+}$ spins \cite{mikhaylovskiy2015ultrafast}.
As a result, the spin correlation function $\hat{C}(\boldsymbol{\delta})$ is impulsively altered along the corresponding crystallographic axis.
In the following, the spin correlation function along this axis will oscillate at the frequency of two-magnon mode which is defined by the energy of the exchange interaction taking twice.
The correlations along the two other axes are affected as well, since different correlations can comprise the same spins. 
Obeying the energy conservation law, spin correlations along the two other crystallographic axes oscillate with the opposite phase, and the total amplitude along all three axes equals zero.
Naturally, such kind of dynamics cannot be comprehended in terms of macrospins $\mathbf{L}$ and $\mathbf{M}$, which do not distinguish anisotropic changes of spin-spin coupling [Suppl. Mater.~\cite{supp_mat}].
Despite this fact, the dynamics manifests itself in the modulation of the symmetric part of the dielectric permittivity which distinguishes the spin correlations along different axes. 
Thus, the spin dynamics is probed by measuring dynamic linear birefringence (for details see Suppl. Mater.~\cite{supp_mat}).
\textcolor{AMK}{We stress that the considered scenario for excitation and detection of the two-magnon mode holds for the isotructural KNiF$_3$~ \cite{bossini2016macrospin,bossini2019laser} as well.
The model suggested in those papers remains valid but represents the next-order effect involving further lowering of the symmetry or strong magnetic anisotropy \cite{fedianin2022selection}.}

In conclusion, we have demonstrated a regime of ultrafast spin dynamics in antiferromagnetic \RMF, where the models considering the antiferromagnetic vector $\mathbf{L}$ and the net magnetization $\mathbf{M}$, fail.
Instead, we propose to describe the regime in terms of the spin correlation function, derive the corresponding equations of motion, and reveal that, unlike the macrospins, the spin correlation function in antiferromagnets does not possess inertia.
As a consequence, a pump pulse with a duration equal to about 1/8th period of the spin wave at the edge of the Brillouin zone\textcolor{AMK}{, i.e. $\approx$30~fs for \RMF,} excites oscillations of spin correlations with the largest amplitude (see Ref.~\cite{fedianin2022selection} for details).
This contrasts with the optimal excitation of spin waves in an antiferromagnet at the center of the Brillouin zone, which simply requires the shortest laser pulses \cite{kimel2015inertia}.
Contrary to models in terms of $\mathbf{M}$ and $\mathbf{L}$, the spin correlation function is capable to describe adequately response to an anisotropic perturbation of short-scale spin-spin exchange coupling by a polarized laser pulse and also reveals the effect of the latter on the optical properties of the medium.
An important outcome of the developed model is the intuitive analytical formula for time-dependent spin dynamics as well as polarization dependences of optical signals in pump-probe experiments, unavailable in earlier reported considerations of spin dynamics at the edge of Brilloun zone~ \cite{zhao2004magnon,bossini2019laser,fedianin2022selection}.
We note that regimes of ultrafast spin dynamics beyond precession of \textbf{L} and \textbf{M} are being also studied by atomistic \cite{Atxitia-PRB2013,Hellsvik-PRB2016} and multiscale simulation \cite{Hinzke-PRB2015}, which, however, focus on strongly-dissipative light-matter interactions and still yield longitudinal and transverse dynamics of \textbf{M} and \textbf{L}.
Our experimental and theoretical findings open up novel perspectives for ultrafast antiferromagnetic spintronics and magnonics governed by laws different from those established for low-energy spin waves.

\textbf{Supplementary Material}. We present a detailed derivation of the equation of motion for spin correlations and the equation for $\varepsilon^{xx}-\varepsilon^{yy}$, and derive the polarization dependences directly from these equations.
We also discuss the key problems one faces when describing the observed dynamics using a N\'eel vecotr \textbf{L}.

\textbf{Acknowledgements}. We are grateful to A.K. Zvezdin, V.N. Gridnev, A.I. Nikitchenko, D. Bossini, H. Hedayat, and P.H.M. van Loosdrecht for fruitful discussions, S. Semin, and C. Berkhout for technical support.
J.\,H.\,M. acknowledges funding by the European Research Council under ERC Grant Agreement No. 856538 (3D-MAGiC), and Horizon Europe project No.~101070290 (NIMFEIA).
The work of T.\,T.\,G. was funded by the European Union's Horizon 2020 research and innovation program under the Marie Sk\l{}odowska-Curie Grant Agreement No.~861300 (COMRAD). 
R.\,M.\,D. acknowledges the support of RSF (Grant No. 22-72-00025).
A.\,M.\,K. acknowledges support from RFBR (Grant No. 19-52-12065).

\textbf{Conflict of Interest}. The authors have no conflicts of interest to disclose. 

The authors declare that this work has been published as a result of peer-to-peer scientific collaboration between researchers. The provided affiliations represent the actual addresses of the authors in agreement with their digital identifier (ORCID) and cannot be considered as a formal collaboration between the aforementioned institutions.

\textbf{Author contributions}. 
$\bold{F.~Formisano}$: Investigation (equal); Writing - review \& editing (supporting). 
$\bold{T.~T.~Gareev}$: Data curation (equal); Formal analysis (equal); Investigation (equal); Methodology (equal); Visualization (equal); Writing - review \& editing (equal). 
$\bold{D.~I.~Khusyainov}$: Data curation (equal); Formal analysis (equal); Investigation (equal); Methodology (equal); Visualization (supporting); Writing - review \& editing (supporting).
$\bold{A.~E.~Fedianin}$: Formal analysis (equal); Investigation (equal); Methodology (equal); Writing - original draft (equal); Writing - review \& editing (equal).
$\bold{R.~M.~Dubrovin}$: Data curation (equal); Formal analysis (equal); Funding acquisition (lead); Investigation (equal); Methodology (equal); Visualization (equal); Writing - original draft (equal); Writing - review \& editing (equal).
$\bold{P.~P.~Syrnikov}$: Resources (equal).
$\bold{D.~Afanasiev}$: Methodology (equal); Supervision (lead); Writing - review \& editing (equal).
$\bold{R.~V.~Pisarev}$: Resources (lead); Writing - review \& editing (equal).
$\bold{A.~M.~Kalashnikova}$: Conceptualization (lead); Funding acquisition (lead); Writing - original draft (lead); Writing - review \& editing (lead).
$\bold{J.~H.~Mentink}$: Conceptualization (lead); Funding acquisition (lead); Methodology (lead); Supervision (lead); Writing - original draft (lead); Writing - review \& editing (lead).
$\bold{A.~V.~Kimel}$: Conceptualization (lead); Funding acquisition (lead); 	Resources (lead); Supervision (lead); Writing - original draft (lead);	Writing - review \& editing (lead).

The data that support the findings of this study are available from the corresponding author upon reasonable request.

\bibliography{bibliography}

\begin{thebibliography}{42}%
\makeatletter
\providecommand \@ifxundefined [1]{%
 \@ifx{#1\undefined}
}%
\providecommand \@ifnum [1]{%
 \ifnum #1\expandafter \@firstoftwo
 \else \expandafter \@secondoftwo
 \fi
}%
\providecommand \@ifx [1]{%
 \ifx #1\expandafter \@firstoftwo
 \else \expandafter \@secondoftwo
 \fi
}%
\providecommand \natexlab [1]{#1}%
\providecommand \enquote  [1]{``#1''}%
\providecommand \bibnamefont  [1]{#1}%
\providecommand \bibfnamefont [1]{#1}%
\providecommand \citenamefont [1]{#1}%
\providecommand \href@noop [0]{\@secondoftwo}%
\providecommand \href [0]{\begingroup \@sanitize@url \@href}%
\providecommand \@href[1]{\@@startlink{#1}\@@href}%
\providecommand \@@href[1]{\endgroup#1\@@endlink}%
\providecommand \@sanitize@url [0]{\catcode `\\12\catcode `\$12\catcode `\&12\catcode `\#12\catcode `\^12\catcode `\_12\catcode `\%12\relax}%
\providecommand \@@startlink[1]{}%
\providecommand \@@endlink[0]{}%
\providecommand \url  [0]{\begingroup\@sanitize@url \@url }%
\providecommand \@url [1]{\endgroup\@href {#1}{\urlprefix }}%
\providecommand \urlprefix  [0]{URL }%
\providecommand \Eprint [0]{\href }%
\providecommand \doibase [0]{https://doi.org/}%
\providecommand \selectlanguage [0]{\@gobble}%
\providecommand \bibinfo  [0]{\@secondoftwo}%
\providecommand \bibfield  [0]{\@secondoftwo}%
\providecommand \translation [1]{[#1]}%
\providecommand \BibitemOpen [0]{}%
\providecommand \bibitemStop [0]{}%
\providecommand \bibitemNoStop [0]{.\EOS\space}%
\providecommand \EOS [0]{\spacefactor3000\relax}%
\providecommand \BibitemShut  [1]{\csname bibitem#1\endcsname}%
\let\auto@bib@innerbib\@empty
\bibitem [{\citenamefont {Pirro}\ \emph {et~al.}(2021)\citenamefont {Pirro}, \citenamefont {Vasyuchka}, \citenamefont {Serga},\ and\ \citenamefont {Hillebrands}}]{pirro2021advances}%
  \BibitemOpen
  \bibfield  {author} {\bibinfo {author} {\bibfnamefont {P.}~\bibnamefont {Pirro}}, \bibinfo {author} {\bibfnamefont {V.~I.}\ \bibnamefont {Vasyuchka}}, \bibinfo {author} {\bibfnamefont {A.~A.}\ \bibnamefont {Serga}},\ and\ \bibinfo {author} {\bibfnamefont {B.}~\bibnamefont {Hillebrands}},\ }\bibfield  {title} {\bibinfo {title} {Advances in coherent magnonics},\ }\href {https://doi.org/10.1038/s41578-021-00332-w} {\bibfield  {journal} {\bibinfo  {journal} {Nat. Rev. Mater.}\ } (\bibinfo {year} {2021})}\BibitemShut {NoStop}%
\bibitem [{\citenamefont {Che}\ \emph {et~al.}(2020)\citenamefont {Che}, \citenamefont {Baumgaertl}, \citenamefont {K{\'u}kol’ov{\'a}}, \citenamefont {Dubs},\ and\ \citenamefont {Grundler}}]{che2020efficient}%
  \BibitemOpen
  \bibfield  {author} {\bibinfo {author} {\bibfnamefont {P.}~\bibnamefont {Che}}, \bibinfo {author} {\bibfnamefont {K.}~\bibnamefont {Baumgaertl}}, \bibinfo {author} {\bibfnamefont {A.}~\bibnamefont {K{\'u}kol’ov{\'a}}}, \bibinfo {author} {\bibfnamefont {C.}~\bibnamefont {Dubs}},\ and\ \bibinfo {author} {\bibfnamefont {D.}~\bibnamefont {Grundler}},\ }\bibfield  {title} {\bibinfo {title} {Efficient wavelength conversion of exchange magnons below 100 nm by magnetic coplanar waveguides},\ }\href {https://doi.org/10.1038/s41467-020-15265-1} {\bibfield  {journal} {\bibinfo  {journal} {Nat. Commun.}\ }\textbf {\bibinfo {volume} {11}},\ \bibinfo {pages} {1145} (\bibinfo {year} {2020})}\BibitemShut {NoStop}%
\bibitem [{\citenamefont {Csaba}\ \emph {et~al.}(2017)\citenamefont {Csaba}, \citenamefont {Papp},\ and\ \citenamefont {Porod}}]{csaba2017perspectives}%
  \BibitemOpen
  \bibfield  {author} {\bibinfo {author} {\bibfnamefont {G.}~\bibnamefont {Csaba}}, \bibinfo {author} {\bibfnamefont {{\'A}.}~\bibnamefont {Papp}},\ and\ \bibinfo {author} {\bibfnamefont {W.}~\bibnamefont {Porod}},\ }\bibfield  {title} {\bibinfo {title} {Perspectives of using spin waves for computing and signal processing},\ }\href {https://doi.org/10.1016/j.physleta.2017.02.042} {\bibfield  {journal} {\bibinfo  {journal} {Phys. Lett. A}\ }\textbf {\bibinfo {volume} {381}},\ \bibinfo {pages} {1471} (\bibinfo {year} {2017})}\BibitemShut {NoStop}%
\bibitem [{\citenamefont {Mentink}\ \emph {et~al.}(2015)\citenamefont {Mentink}, \citenamefont {Balzer},\ and\ \citenamefont {Eckstein}}]{mentink2015ultrafast}%
  \BibitemOpen
  \bibfield  {author} {\bibinfo {author} {\bibfnamefont {J.}~\bibnamefont {Mentink}}, \bibinfo {author} {\bibfnamefont {K.}~\bibnamefont {Balzer}},\ and\ \bibinfo {author} {\bibfnamefont {M.}~\bibnamefont {Eckstein}},\ }\bibfield  {title} {\bibinfo {title} {Ultrafast and reversible control of the exchange interaction in {M}ott insulators},\ }\href {https://doi.org/10.1038/ncomms7708} {\bibfield  {journal} {\bibinfo  {journal} {Nat. Commun.}\ }\textbf {\bibinfo {volume} {6}},\ \bibinfo {pages} {6708} (\bibinfo {year} {2015})}\BibitemShut {NoStop}%
\bibitem [{\citenamefont {Sandweg}\ \emph {et~al.}(2011)\citenamefont {Sandweg}, \citenamefont {Kajiwara}, \citenamefont {Chumak}, \citenamefont {Serga}, \citenamefont {Vasyuchka}, \citenamefont {Jungfleisch}, \citenamefont {Saitoh},\ and\ \citenamefont {Hillebrands}}]{sandweg2011spin}%
  \BibitemOpen
  \bibfield  {author} {\bibinfo {author} {\bibfnamefont {C.~W.}\ \bibnamefont {Sandweg}}, \bibinfo {author} {\bibfnamefont {Y.}~\bibnamefont {Kajiwara}}, \bibinfo {author} {\bibfnamefont {A.~V.}\ \bibnamefont {Chumak}}, \bibinfo {author} {\bibfnamefont {A.~A.}\ \bibnamefont {Serga}}, \bibinfo {author} {\bibfnamefont {V.~I.}\ \bibnamefont {Vasyuchka}}, \bibinfo {author} {\bibfnamefont {M.~B.}\ \bibnamefont {Jungfleisch}}, \bibinfo {author} {\bibfnamefont {E.}~\bibnamefont {Saitoh}},\ and\ \bibinfo {author} {\bibfnamefont {B.}~\bibnamefont {Hillebrands}},\ }\bibfield  {title} {\bibinfo {title} {{S}pin {P}umping by {P}arametrically {E}xcited {E}xchange {M}agnons},\ }\href {https://doi.org/10.1103/PhysRevLett.106.216601} {\bibfield  {journal} {\bibinfo  {journal} {Phys. Rev. Lett.}\ }\textbf {\bibinfo {volume} {106}},\ \bibinfo {pages} {216601} (\bibinfo {year} {2011})}\BibitemShut {NoStop}%
\bibitem [{\citenamefont {Barman}\ \emph {et~al.}(2021)\citenamefont {Barman}, \citenamefont {Gubbiotti}, \citenamefont {Ladak}, \citenamefont {Adeyeye}, \citenamefont {Krawczyk}, \citenamefont {Gräfe}, \citenamefont {Adelmann}, \citenamefont {Cotofana}, \citenamefont {Naeemi}, \citenamefont {Vasyuchka}, \citenamefont {Hillebrands}, \citenamefont {Nikitov}, \citenamefont {Yu}, \citenamefont {Grundler}, \citenamefont {Sadovnikov}, \citenamefont {Grachev}, \citenamefont {Sheshukova}, \citenamefont {Duquesne}, \citenamefont {Marangolo}, \citenamefont {Csaba}, \citenamefont {Porod}, \citenamefont {Demidov}, \citenamefont {Urazhdin}, \citenamefont {Demokritov}, \citenamefont {Albisetti}, \citenamefont {Petti}, \citenamefont {Bertacco}, \citenamefont {Schultheiss}, \citenamefont {Kruglyak}, \citenamefont {Poimanov}, \citenamefont {Sahoo}, \citenamefont {Sinha}, \citenamefont {Yang}, \citenamefont {Münzenberg}, \citenamefont {Moriyama}, \citenamefont {Mizukami}, \citenamefont {Landeros}, \citenamefont {Gallardo},
  \citenamefont {Carlotti}, \citenamefont {Kim}, \citenamefont {Stamps}, \citenamefont {Camley}, \citenamefont {Rana}, \citenamefont {Otani}, \citenamefont {Yu}, \citenamefont {Yu}, \citenamefont {Bauer}, \citenamefont {Back}, \citenamefont {Uhrig}, \citenamefont {Dobrovolskiy}, \citenamefont {Budinska}, \citenamefont {Qin}, \citenamefont {van Dijken}, \citenamefont {Chumak}, \citenamefont {Khitun}, \citenamefont {Nikonov}, \citenamefont {Young}, \citenamefont {Zingsem},\ and\ \citenamefont {Winklhofer}}]{barman2021roadmap}%
  \BibitemOpen
  \bibfield  {author} {\bibinfo {author} {\bibfnamefont {A.}~\bibnamefont {Barman}}, \bibinfo {author} {\bibfnamefont {G.}~\bibnamefont {Gubbiotti}}, \bibinfo {author} {\bibfnamefont {S.}~\bibnamefont {Ladak}}, \bibinfo {author} {\bibfnamefont {A.~O.}\ \bibnamefont {Adeyeye}}, \bibinfo {author} {\bibfnamefont {M.}~\bibnamefont {Krawczyk}}, \bibinfo {author} {\bibfnamefont {J.}~\bibnamefont {Gräfe}}, \bibinfo {author} {\bibfnamefont {C.}~\bibnamefont {Adelmann}}, \bibinfo {author} {\bibfnamefont {S.}~\bibnamefont {Cotofana}}, \bibinfo {author} {\bibfnamefont {A.}~\bibnamefont {Naeemi}}, \bibinfo {author} {\bibfnamefont {V.~I.}\ \bibnamefont {Vasyuchka}}, \bibinfo {author} {\bibfnamefont {B.}~\bibnamefont {Hillebrands}}, \bibinfo {author} {\bibfnamefont {S.~A.}\ \bibnamefont {Nikitov}}, \bibinfo {author} {\bibfnamefont {H.}~\bibnamefont {Yu}}, \bibinfo {author} {\bibfnamefont {D.}~\bibnamefont {Grundler}}, \bibinfo {author} {\bibfnamefont {A.~V.}\ \bibnamefont {Sadovnikov}}, \bibinfo {author} {\bibfnamefont
  {A.~A.}\ \bibnamefont {Grachev}}, \bibinfo {author} {\bibfnamefont {S.~E.}\ \bibnamefont {Sheshukova}}, \bibinfo {author} {\bibfnamefont {J.-Y.}\ \bibnamefont {Duquesne}}, \bibinfo {author} {\bibfnamefont {M.}~\bibnamefont {Marangolo}}, \bibinfo {author} {\bibfnamefont {G.}~\bibnamefont {Csaba}}, \bibinfo {author} {\bibfnamefont {W.}~\bibnamefont {Porod}}, \bibinfo {author} {\bibfnamefont {V.~E.}\ \bibnamefont {Demidov}}, \bibinfo {author} {\bibfnamefont {S.}~\bibnamefont {Urazhdin}}, \bibinfo {author} {\bibfnamefont {S.~O.}\ \bibnamefont {Demokritov}}, \bibinfo {author} {\bibfnamefont {E.}~\bibnamefont {Albisetti}}, \bibinfo {author} {\bibfnamefont {D.}~\bibnamefont {Petti}}, \bibinfo {author} {\bibfnamefont {R.}~\bibnamefont {Bertacco}}, \bibinfo {author} {\bibfnamefont {H.}~\bibnamefont {Schultheiss}}, \bibinfo {author} {\bibfnamefont {V.~V.}\ \bibnamefont {Kruglyak}}, \bibinfo {author} {\bibfnamefont {V.~D.}\ \bibnamefont {Poimanov}}, \bibinfo {author} {\bibfnamefont {S.}~\bibnamefont {Sahoo}}, \bibinfo
  {author} {\bibfnamefont {J.}~\bibnamefont {Sinha}}, \bibinfo {author} {\bibfnamefont {H.}~\bibnamefont {Yang}}, \bibinfo {author} {\bibfnamefont {M.}~\bibnamefont {Münzenberg}}, \bibinfo {author} {\bibfnamefont {T.}~\bibnamefont {Moriyama}}, \bibinfo {author} {\bibfnamefont {S.}~\bibnamefont {Mizukami}}, \bibinfo {author} {\bibfnamefont {P.}~\bibnamefont {Landeros}}, \bibinfo {author} {\bibfnamefont {R.~A.}\ \bibnamefont {Gallardo}}, \bibinfo {author} {\bibfnamefont {G.}~\bibnamefont {Carlotti}}, \bibinfo {author} {\bibfnamefont {J.-V.}\ \bibnamefont {Kim}}, \bibinfo {author} {\bibfnamefont {R.~L.}\ \bibnamefont {Stamps}}, \bibinfo {author} {\bibfnamefont {R.~E.}\ \bibnamefont {Camley}}, \bibinfo {author} {\bibfnamefont {B.}~\bibnamefont {Rana}}, \bibinfo {author} {\bibfnamefont {Y.}~\bibnamefont {Otani}}, \bibinfo {author} {\bibfnamefont {W.}~\bibnamefont {Yu}}, \bibinfo {author} {\bibfnamefont {T.}~\bibnamefont {Yu}}, \bibinfo {author} {\bibfnamefont {G.~E.~W.}\ \bibnamefont {Bauer}}, \bibinfo {author}
  {\bibfnamefont {C.}~\bibnamefont {Back}}, \bibinfo {author} {\bibfnamefont {G.~S.}\ \bibnamefont {Uhrig}}, \bibinfo {author} {\bibfnamefont {O.~V.}\ \bibnamefont {Dobrovolskiy}}, \bibinfo {author} {\bibfnamefont {B.}~\bibnamefont {Budinska}}, \bibinfo {author} {\bibfnamefont {H.}~\bibnamefont {Qin}}, \bibinfo {author} {\bibfnamefont {S.}~\bibnamefont {van Dijken}}, \bibinfo {author} {\bibfnamefont {A.~V.}\ \bibnamefont {Chumak}}, \bibinfo {author} {\bibfnamefont {A.}~\bibnamefont {Khitun}}, \bibinfo {author} {\bibfnamefont {D.~E.}\ \bibnamefont {Nikonov}}, \bibinfo {author} {\bibfnamefont {I.~A.}\ \bibnamefont {Young}}, \bibinfo {author} {\bibfnamefont {B.~W.}\ \bibnamefont {Zingsem}},\ and\ \bibinfo {author} {\bibfnamefont {M.}~\bibnamefont {Winklhofer}},\ }\bibfield  {title} {\bibinfo {title} {{T}he 2021 {M}agnonics {R}oadmap},\ }\href {https://doi.org/10.1088/1361-648X/abec1a} {\bibfield  {journal} {\bibinfo  {journal} {J. Phys. Condens. Matter}\ }\textbf {\bibinfo {volume} {33}},\ \bibinfo {pages}
  {413001} (\bibinfo {year} {2021})}\BibitemShut {NoStop}%
\bibitem [{\citenamefont {Fukami}\ \emph {et~al.}(2020)\citenamefont {Fukami}, \citenamefont {Lorenz},\ and\ \citenamefont {Gomonay}}]{fukami2020antiferromagnetic}%
  \BibitemOpen
  \bibfield  {author} {\bibinfo {author} {\bibfnamefont {S.}~\bibnamefont {Fukami}}, \bibinfo {author} {\bibfnamefont {V.~O.}\ \bibnamefont {Lorenz}},\ and\ \bibinfo {author} {\bibfnamefont {O.}~\bibnamefont {Gomonay}},\ }\bibfield  {title} {\bibinfo {title} {Antiferromagnetic spintronics},\ }\href {https://doi.org/10.1063/5.0023614} {\bibfield  {journal} {\bibinfo  {journal} {J. Appl. Phys.}\ }\textbf {\bibinfo {volume} {128}},\ \bibinfo {pages} {070401} (\bibinfo {year} {2020})}\BibitemShut {NoStop}%
\bibitem [{\citenamefont {N{\v{e}}mec}\ \emph {et~al.}(2018)\citenamefont {N{\v{e}}mec}, \citenamefont {Fiebig}, \citenamefont {Kampfrath},\ and\ \citenamefont {Kimel}}]{nemec2018antiferromagnetic}%
  \BibitemOpen
  \bibfield  {author} {\bibinfo {author} {\bibfnamefont {P.}~\bibnamefont {N{\v{e}}mec}}, \bibinfo {author} {\bibfnamefont {M.}~\bibnamefont {Fiebig}}, \bibinfo {author} {\bibfnamefont {T.}~\bibnamefont {Kampfrath}},\ and\ \bibinfo {author} {\bibfnamefont {A.~V.}\ \bibnamefont {Kimel}},\ }\bibfield  {title} {\bibinfo {title} {Antiferromagnetic opto-spintronics},\ }\href {https://doi.org/10.1038/s41567-018-0051-x} {\bibfield  {journal} {\bibinfo  {journal} {Nat. Phys.}\ }\textbf {\bibinfo {volume} {14}},\ \bibinfo {pages} {229} (\bibinfo {year} {2018})}\BibitemShut {NoStop}%
\bibitem [{\citenamefont {Jungwirth}\ \emph {et~al.}(2018)\citenamefont {Jungwirth}, \citenamefont {Sinova}, \citenamefont {Manchon}, \citenamefont {Marti}, \citenamefont {Wunderlich},\ and\ \citenamefont {Felser}}]{jungwirth2018multiple}%
  \BibitemOpen
  \bibfield  {author} {\bibinfo {author} {\bibfnamefont {T.}~\bibnamefont {Jungwirth}}, \bibinfo {author} {\bibfnamefont {J.}~\bibnamefont {Sinova}}, \bibinfo {author} {\bibfnamefont {A.}~\bibnamefont {Manchon}}, \bibinfo {author} {\bibfnamefont {X.}~\bibnamefont {Marti}}, \bibinfo {author} {\bibfnamefont {J.}~\bibnamefont {Wunderlich}},\ and\ \bibinfo {author} {\bibfnamefont {C.}~\bibnamefont {Felser}},\ }\bibfield  {title} {\bibinfo {title} {The multiple directions of antiferromagnetic spintronics},\ }\href {https://doi.org/10.1038/s41567-018-0063-6} {\bibfield  {journal} {\bibinfo  {journal} {Nat. Phys.}\ }\textbf {\bibinfo {volume} {14}},\ \bibinfo {pages} {200} (\bibinfo {year} {2018})}\BibitemShut {NoStop}%
\bibitem [{\citenamefont {Baltz}\ \emph {et~al.}(2018)\citenamefont {Baltz}, \citenamefont {Manchon}, \citenamefont {Tsoi}, \citenamefont {Moriyama}, \citenamefont {Ono},\ and\ \citenamefont {Tserkovnyak}}]{baltz2018antiferromagnetic}%
  \BibitemOpen
  \bibfield  {author} {\bibinfo {author} {\bibfnamefont {V.}~\bibnamefont {Baltz}}, \bibinfo {author} {\bibfnamefont {A.}~\bibnamefont {Manchon}}, \bibinfo {author} {\bibfnamefont {M.}~\bibnamefont {Tsoi}}, \bibinfo {author} {\bibfnamefont {T.}~\bibnamefont {Moriyama}}, \bibinfo {author} {\bibfnamefont {T.}~\bibnamefont {Ono}},\ and\ \bibinfo {author} {\bibfnamefont {Y.}~\bibnamefont {Tserkovnyak}},\ }\bibfield  {title} {\bibinfo {title} {Antiferromagnetic spintronics},\ }\href {https://doi.org/10.1103/RevModPhys.90.015005} {\bibfield  {journal} {\bibinfo  {journal} {Rev. Mod. Phys.}\ }\textbf {\bibinfo {volume} {90}},\ \bibinfo {pages} {015005} (\bibinfo {year} {2018})}\BibitemShut {NoStop}%
\bibitem [{\citenamefont {Jungwirth}\ \emph {et~al.}(2016)\citenamefont {Jungwirth}, \citenamefont {Marti}, \citenamefont {Wadley},\ and\ \citenamefont {Wunderlich}}]{jungwirth2016antiferromagnetic}%
  \BibitemOpen
  \bibfield  {author} {\bibinfo {author} {\bibfnamefont {T.}~\bibnamefont {Jungwirth}}, \bibinfo {author} {\bibfnamefont {X.}~\bibnamefont {Marti}}, \bibinfo {author} {\bibfnamefont {P.}~\bibnamefont {Wadley}},\ and\ \bibinfo {author} {\bibfnamefont {J.}~\bibnamefont {Wunderlich}},\ }\bibfield  {title} {\bibinfo {title} {Antiferromagnetic spintronics},\ }\href {https://doi.org/10.1038/nnano.2016.18} {\bibfield  {journal} {\bibinfo  {journal} {Nat. Nanotechnol.}\ }\textbf {\bibinfo {volume} {11}},\ \bibinfo {pages} {231} (\bibinfo {year} {2016})}\BibitemShut {NoStop}%
\bibitem [{\citenamefont {Kalashnikova}\ \emph {et~al.}(2007)\citenamefont {Kalashnikova}, \citenamefont {Kimel}, \citenamefont {Pisarev}, \citenamefont {Gridnev}, \citenamefont {Kirilyuk},\ and\ \citenamefont {Rasing}}]{kalashnikova2007impulsive}%
  \BibitemOpen
  \bibfield  {author} {\bibinfo {author} {\bibfnamefont {A.~M.}\ \bibnamefont {Kalashnikova}}, \bibinfo {author} {\bibfnamefont {A.~V.}\ \bibnamefont {Kimel}}, \bibinfo {author} {\bibfnamefont {R.~V.}\ \bibnamefont {Pisarev}}, \bibinfo {author} {\bibfnamefont {V.~N.}\ \bibnamefont {Gridnev}}, \bibinfo {author} {\bibfnamefont {A.}~\bibnamefont {Kirilyuk}},\ and\ \bibinfo {author} {\bibfnamefont {T.}~\bibnamefont {Rasing}},\ }\bibfield  {title} {\bibinfo {title} {{I}mpulsive {G}eneration of {C}oherent {M}agnons by {L}inearly {P}olarized {L}ight in the {E}asy-{P}lane {A}ntiferromagnet $\mathrm{FeBO}_{3}$},\ }\href {https://doi.org/10.1103/PhysRevLett.99.167205} {\bibfield  {journal} {\bibinfo  {journal} {Phys. Rev. Lett.}\ }\textbf {\bibinfo {volume} {99}},\ \bibinfo {pages} {167205} (\bibinfo {year} {2007})}\BibitemShut {NoStop}%
\bibitem [{\citenamefont {Gridnev}(2008)}]{gridnev2008phenomenological}%
  \BibitemOpen
  \bibfield  {author} {\bibinfo {author} {\bibfnamefont {V.~N.}\ \bibnamefont {Gridnev}},\ }\bibfield  {title} {\bibinfo {title} {Phenomenological theory for coherent magnon generation through impulsive stimulated {R}aman scattering},\ }\href {https://doi.org/10.1103/PhysRevB.77.094426} {\bibfield  {journal} {\bibinfo  {journal} {Phys. Rev. B}\ }\textbf {\bibinfo {volume} {77}},\ \bibinfo {pages} {094426} (\bibinfo {year} {2008})}\BibitemShut {NoStop}%
\bibitem [{\citenamefont {Satoh}\ \emph {et~al.}(2015)\citenamefont {Satoh}, \citenamefont {Iida}, \citenamefont {Higuchi}, \citenamefont {Fiebig},\ and\ \citenamefont {Shimura}}]{satoh2015writing}%
  \BibitemOpen
  \bibfield  {author} {\bibinfo {author} {\bibfnamefont {T.}~\bibnamefont {Satoh}}, \bibinfo {author} {\bibfnamefont {R.}~\bibnamefont {Iida}}, \bibinfo {author} {\bibfnamefont {T.}~\bibnamefont {Higuchi}}, \bibinfo {author} {\bibfnamefont {M.}~\bibnamefont {Fiebig}},\ and\ \bibinfo {author} {\bibfnamefont {T.}~\bibnamefont {Shimura}},\ }\bibfield  {title} {\bibinfo {title} {Writing and reading of an arbitrary optical polarization state in an antiferromagnet},\ }\href {https://doi.org/10.1038/nphoton.2014.273} {\bibfield  {journal} {\bibinfo  {journal} {Nature Photon.}\ }\textbf {\bibinfo {volume} {9}},\ \bibinfo {pages} {25} (\bibinfo {year} {2015})}\BibitemShut {NoStop}%
\bibitem [{\citenamefont {Tzschaschel}\ \emph {et~al.}(2017)\citenamefont {Tzschaschel}, \citenamefont {Otani}, \citenamefont {Iida}, \citenamefont {Shimura}, \citenamefont {Ueda}, \citenamefont {G\"unther}, \citenamefont {Fiebig},\ and\ \citenamefont {Satoh}}]{tzschaschel2017ultrafast}%
  \BibitemOpen
  \bibfield  {author} {\bibinfo {author} {\bibfnamefont {C.}~\bibnamefont {Tzschaschel}}, \bibinfo {author} {\bibfnamefont {K.}~\bibnamefont {Otani}}, \bibinfo {author} {\bibfnamefont {R.}~\bibnamefont {Iida}}, \bibinfo {author} {\bibfnamefont {T.}~\bibnamefont {Shimura}}, \bibinfo {author} {\bibfnamefont {H.}~\bibnamefont {Ueda}}, \bibinfo {author} {\bibfnamefont {S.}~\bibnamefont {G\"unther}}, \bibinfo {author} {\bibfnamefont {M.}~\bibnamefont {Fiebig}},\ and\ \bibinfo {author} {\bibfnamefont {T.}~\bibnamefont {Satoh}},\ }\bibfield  {title} {\bibinfo {title} {Ultrafast optical excitation of coherent magnons in antiferromagnetic nio},\ }\href {https://doi.org/10.1103/PhysRevB.95.174407} {\bibfield  {journal} {\bibinfo  {journal} {Phys. Rev. B}\ }\textbf {\bibinfo {volume} {95}},\ \bibinfo {pages} {174407} (\bibinfo {year} {2017})}\BibitemShut {NoStop}%
\bibitem [{\citenamefont {Satoh}\ \emph {et~al.}(2012)\citenamefont {Satoh}, \citenamefont {Terui}, \citenamefont {Moriya}, \citenamefont {Ivanov}, \citenamefont {Ando}, \citenamefont {Saitoh}, \citenamefont {Shimura},\ and\ \citenamefont {Kazuo}}]{satoh2012spinwaves}%
  \BibitemOpen
  \bibfield  {author} {\bibinfo {author} {\bibfnamefont {T.}~\bibnamefont {Satoh}}, \bibinfo {author} {\bibfnamefont {Y.}~\bibnamefont {Terui}}, \bibinfo {author} {\bibfnamefont {R.}~\bibnamefont {Moriya}}, \bibinfo {author} {\bibfnamefont {B.~A.}\ \bibnamefont {Ivanov}}, \bibinfo {author} {\bibfnamefont {K.}~\bibnamefont {Ando}}, \bibinfo {author} {\bibfnamefont {E.}~\bibnamefont {Saitoh}}, \bibinfo {author} {\bibfnamefont {T.}~\bibnamefont {Shimura}},\ and\ \bibinfo {author} {\bibfnamefont {K.}~\bibnamefont {Kazuo}},\ }\bibfield  {title} {\bibinfo {title} {Directional control of spin-wave emission by spatially shaped light},\ }\href {https://doi.org/10.1038/nphoton.2012.218} {\bibfield  {journal} {\bibinfo  {journal} {Nature Photon.}\ }\textbf {\bibinfo {volume} {6}},\ \bibinfo {pages} {662} (\bibinfo {year} {2012})}\BibitemShut {NoStop}%
\bibitem [{\citenamefont {Hortensius}\ \emph {et~al.}(2021)\citenamefont {Hortensius}, \citenamefont {Afanasiev}, \citenamefont {Matthiesen}, \citenamefont {Leenders}, \citenamefont {Citro}, \citenamefont {Kimel}, \citenamefont {Mikhaylovskiy}, \citenamefont {Ivanov},\ and\ \citenamefont {Caviglia}}]{afanasiev2021spinwaves}%
  \BibitemOpen
  \bibfield  {author} {\bibinfo {author} {\bibfnamefont {J.~R.}\ \bibnamefont {Hortensius}}, \bibinfo {author} {\bibfnamefont {D.}~\bibnamefont {Afanasiev}}, \bibinfo {author} {\bibfnamefont {M.}~\bibnamefont {Matthiesen}}, \bibinfo {author} {\bibfnamefont {R.}~\bibnamefont {Leenders}}, \bibinfo {author} {\bibfnamefont {R.}~\bibnamefont {Citro}}, \bibinfo {author} {\bibfnamefont {A.~V.}\ \bibnamefont {Kimel}}, \bibinfo {author} {\bibfnamefont {R.~V.}\ \bibnamefont {Mikhaylovskiy}}, \bibinfo {author} {\bibfnamefont {B.~A.}\ \bibnamefont {Ivanov}},\ and\ \bibinfo {author} {\bibfnamefont {A.~D.}\ \bibnamefont {Caviglia}},\ }\bibfield  {title} {\bibinfo {title} {Coherent spin-wave transport in an antiferromagnet},\ }\href {https://doi.org/10.1038/s41567-021-01290-4} {\bibfield  {journal} {\bibinfo  {journal} {Nat. Phys.}\ }\textbf {\bibinfo {volume} {17}},\ \bibinfo {pages} {1001–} (\bibinfo {year} {2021})}\BibitemShut {NoStop}%
\bibitem [{\citenamefont {Kotyuzhanskii}\ and\ \citenamefont {Prozorova}(1981)}]{kotyuzhanskii1981parametric}%
  \BibitemOpen
  \bibfield  {author} {\bibinfo {author} {\bibfnamefont {B.~Y.}\ \bibnamefont {Kotyuzhanskii}}\ and\ \bibinfo {author} {\bibfnamefont {L.~A.}\ \bibnamefont {Prozorova}},\ }\bibfield  {title} {\bibinfo {title} {Parametric excitation of spin waves in the antiferromagnet $\mathrm{FeBO}_{3}$},\ }\href {http://www.jetp.ras.ru/cgi-bin/dn/e_054_05_1013.pdf} {\bibfield  {journal} {\bibinfo  {journal} {Zh. Eksp. Teor. Fiz}\ }\textbf {\bibinfo {volume} {81}},\ \bibinfo {pages} {1913} (\bibinfo {year} {1981})}\BibitemShut {NoStop}%
\bibitem [{\citenamefont {Fleury}\ and\ \citenamefont {Loudon}(1968)}]{fleury1968scattering}%
  \BibitemOpen
  \bibfield  {author} {\bibinfo {author} {\bibfnamefont {P.~A.}\ \bibnamefont {Fleury}}\ and\ \bibinfo {author} {\bibfnamefont {R.}~\bibnamefont {Loudon}},\ }\bibfield  {title} {\bibinfo {title} {{S}cattering of {L}ight by {O}ne- and {T}wo-{M}agnon {E}citations},\ }\href {https://doi.org/10.1103/PhysRev.166.514} {\bibfield  {journal} {\bibinfo  {journal} {Phys. Rev.}\ }\textbf {\bibinfo {volume} {166}},\ \bibinfo {pages} {514} (\bibinfo {year} {1968})}\BibitemShut {NoStop}%
\bibitem [{\citenamefont {Zhao}\ \emph {et~al.}(2004)\citenamefont {Zhao}, \citenamefont {Bragas}, \citenamefont {Lockwood},\ and\ \citenamefont {Merlin}}]{zhao2004magnon}%
  \BibitemOpen
  \bibfield  {author} {\bibinfo {author} {\bibfnamefont {J.}~\bibnamefont {Zhao}}, \bibinfo {author} {\bibfnamefont {A.~V.}\ \bibnamefont {Bragas}}, \bibinfo {author} {\bibfnamefont {D.~J.}\ \bibnamefont {Lockwood}},\ and\ \bibinfo {author} {\bibfnamefont {R.}~\bibnamefont {Merlin}},\ }\bibfield  {title} {\bibinfo {title} {{M}agnon {S}queezing in an {A}ntiferromagnet: {R}educing the {S}pin {N}oise below the {S}tandard {Q}uantum {L}imit},\ }\href {https://doi.org/10.1103/PhysRevLett.93.107203} {\bibfield  {journal} {\bibinfo  {journal} {Phys. Rev. Lett.}\ }\textbf {\bibinfo {volume} {93}},\ \bibinfo {pages} {107203} (\bibinfo {year} {2004})}\BibitemShut {NoStop}%
\bibitem [{\citenamefont {Bossini}\ \emph {et~al.}(2016)\citenamefont {Bossini}, \citenamefont {Dal~Conte}, \citenamefont {Hashimoto}, \citenamefont {Secchi}, \citenamefont {Pisarev}, \citenamefont {Rasing}, \citenamefont {Cerullo},\ and\ \citenamefont {Kimel}}]{bossini2016macrospin}%
  \BibitemOpen
  \bibfield  {author} {\bibinfo {author} {\bibfnamefont {D.}~\bibnamefont {Bossini}}, \bibinfo {author} {\bibfnamefont {S.}~\bibnamefont {Dal~Conte}}, \bibinfo {author} {\bibfnamefont {Y.}~\bibnamefont {Hashimoto}}, \bibinfo {author} {\bibfnamefont {A.}~\bibnamefont {Secchi}}, \bibinfo {author} {\bibfnamefont {R.~V.}\ \bibnamefont {Pisarev}}, \bibinfo {author} {\bibfnamefont {T.}~\bibnamefont {Rasing}}, \bibinfo {author} {\bibfnamefont {G.}~\bibnamefont {Cerullo}},\ and\ \bibinfo {author} {\bibfnamefont {A.~V.}\ \bibnamefont {Kimel}},\ }\bibfield  {title} {\bibinfo {title} {Macrospin dynamics in antiferromagnets triggered by sub-20 femtosecond injection of nanomagnons},\ }\href {https://doi.org/10.1038/ncomms10645} {\bibfield  {journal} {\bibinfo  {journal} {Nat. Commun.}\ }\textbf {\bibinfo {volume} {7}},\ \bibinfo {pages} {10645} (\bibinfo {year} {2016})}\BibitemShut {NoStop}%
\bibitem [{\citenamefont {Kimel}\ \emph {et~al.}(2020)\citenamefont {Kimel}, \citenamefont {Kalashnikova}, \citenamefont {Pogrebna},\ and\ \citenamefont {Zvezdin}}]{kimel2020fundamentals}%
  \BibitemOpen
  \bibfield  {author} {\bibinfo {author} {\bibfnamefont {A.~V.}\ \bibnamefont {Kimel}}, \bibinfo {author} {\bibfnamefont {A.~M.}\ \bibnamefont {Kalashnikova}}, \bibinfo {author} {\bibfnamefont {A.}~\bibnamefont {Pogrebna}},\ and\ \bibinfo {author} {\bibfnamefont {A.~K.}\ \bibnamefont {Zvezdin}},\ }\bibfield  {title} {\bibinfo {title} {Fundamentals and perspectives of ultrafast photoferroic recording},\ }\href {https://doi.org/10.1016/j.physrep.2020.01.004} {\bibfield  {journal} {\bibinfo  {journal} {Phys. Rep.}\ }\textbf {\bibinfo {volume} {852}},\ \bibinfo {pages} {1} (\bibinfo {year} {2020})}\BibitemShut {NoStop}%
\bibitem [{\citenamefont {Teaney}\ \emph {et~al.}(1962)\citenamefont {Teaney}, \citenamefont {Freiser},\ and\ \citenamefont {Stevenson}}]{teaney1962discovery}%
  \BibitemOpen
  \bibfield  {author} {\bibinfo {author} {\bibfnamefont {D.~T.}\ \bibnamefont {Teaney}}, \bibinfo {author} {\bibfnamefont {M.~J.}\ \bibnamefont {Freiser}},\ and\ \bibinfo {author} {\bibfnamefont {R.~W.~H.}\ \bibnamefont {Stevenson}},\ }\bibfield  {title} {\bibinfo {title} {{D}iscovery of a {S}imple {C}ubic {A}ntiferromagnet: {A}ntiferromagnetic {R}esonance in $\mathrm{RbMnF}_{3}$},\ }\href {https://doi.org/10.1103/PhysRevLett.9.212} {\bibfield  {journal} {\bibinfo  {journal} {Phys. Rev. Lett.}\ }\textbf {\bibinfo {volume} {9}},\ \bibinfo {pages} {212} (\bibinfo {year} {1962})}\BibitemShut {NoStop}%
\bibitem [{\citenamefont {Freiser}\ \emph {et~al.}(1963)\citenamefont {Freiser}, \citenamefont {Seiden},\ and\ \citenamefont {Teaney}}]{freiser1963field}%
  \BibitemOpen
  \bibfield  {author} {\bibinfo {author} {\bibfnamefont {M.~J.}\ \bibnamefont {Freiser}}, \bibinfo {author} {\bibfnamefont {P.~E.}\ \bibnamefont {Seiden}},\ and\ \bibinfo {author} {\bibfnamefont {D.~T.}\ \bibnamefont {Teaney}},\ }\bibfield  {title} {\bibinfo {title} {{F}ield-{I}ndependent {L}ongitudinal {A}ntiferromagnetic {R}esonance},\ }\href {https://doi.org/10.1103/PhysRevLett.10.293} {\bibfield  {journal} {\bibinfo  {journal} {Phys. Rev. Lett.}\ }\textbf {\bibinfo {volume} {10}},\ \bibinfo {pages} {293} (\bibinfo {year} {1963})}\BibitemShut {NoStop}%
\bibitem [{\citenamefont {L\'opez~Ortiz}\ \emph {et~al.}(2014)\citenamefont {L\'opez~Ortiz}, \citenamefont {Fonseca~Guerra}, \citenamefont {Machado},\ and\ \citenamefont {Rezende}}]{lopez2014magnetic}%
  \BibitemOpen
  \bibfield  {author} {\bibinfo {author} {\bibfnamefont {J.~C.}\ \bibnamefont {L\'opez~Ortiz}}, \bibinfo {author} {\bibfnamefont {G.~A.}\ \bibnamefont {Fonseca~Guerra}}, \bibinfo {author} {\bibfnamefont {F.~L.~A.}\ \bibnamefont {Machado}},\ and\ \bibinfo {author} {\bibfnamefont {S.~M.}\ \bibnamefont {Rezende}},\ }\bibfield  {title} {\bibinfo {title} {Magnetic anisotropy of antiferromagnetic $\mathrm{RbMnF}_{3}$},\ }\href {https://doi.org/10.1103/PhysRevB.90.054402} {\bibfield  {journal} {\bibinfo  {journal} {Phys. Rev. B}\ }\textbf {\bibinfo {volume} {90}},\ \bibinfo {pages} {054402} (\bibinfo {year} {2014})}\BibitemShut {NoStop}%
\bibitem [{\citenamefont {Fleury}(1968)}]{fleury1968evidence}%
  \BibitemOpen
  \bibfield  {author} {\bibinfo {author} {\bibfnamefont {P.~A.}\ \bibnamefont {Fleury}},\ }\bibfield  {title} {\bibinfo {title} {Evidence for {M}agnon-{M}agnon {I}nteractions in $\mathrm{RbMnF}_{3}$},\ }\href {https://doi.org/10.1103/PhysRevLett.21.151} {\bibfield  {journal} {\bibinfo  {journal} {Phys. Rev. Lett.}\ }\textbf {\bibinfo {volume} {21}},\ \bibinfo {pages} {151} (\bibinfo {year} {1968})}\BibitemShut {NoStop}%
\bibitem [{\citenamefont {Elliott}\ and\ \citenamefont {Thorpe}(1969)}]{elliott1969effects}%
  \BibitemOpen
  \bibfield  {author} {\bibinfo {author} {\bibfnamefont {R.~J.}\ \bibnamefont {Elliott}}\ and\ \bibinfo {author} {\bibfnamefont {M.~F.}\ \bibnamefont {Thorpe}},\ }\bibfield  {title} {\bibinfo {title} {The effects of magnon-magnon interaction on the two-magnon spectra of antiferromagnets},\ }\href {https://doi.org/10.1088/0022-3719/2/9/312} {\bibfield  {journal} {\bibinfo  {journal} {J. Phys. C: Solid State Phys.}\ }\textbf {\bibinfo {volume} {2}},\ \bibinfo {pages} {1630} (\bibinfo {year} {1969})}\BibitemShut {NoStop}%
\bibitem [{\citenamefont {Fleury}(1970)}]{fleury1970temperature}%
  \BibitemOpen
  \bibfield  {author} {\bibinfo {author} {\bibfnamefont {P.~A.}\ \bibnamefont {Fleury}},\ }\bibfield  {title} {\bibinfo {title} {{T}emperature {D}ependence of {M}agnon-{P}air {M}odes in {A}ntiferromagnets and {P}aramagnets},\ }\href {https://doi.org/10.1063/1.1659002} {\bibfield  {journal} {\bibinfo  {journal} {J. Appl. Phys.}\ }\textbf {\bibinfo {volume} {41}},\ \bibinfo {pages} {886} (\bibinfo {year} {1970})}\BibitemShut {NoStop}%
\bibitem [{\citenamefont {Barocchi}\ \emph {et~al.}(1978)\citenamefont {Barocchi}, \citenamefont {Mazzinghi}, \citenamefont {Tognetti},\ and\ \citenamefont {Zoppi}}]{barocchi1978determination}%
  \BibitemOpen
  \bibfield  {author} {\bibinfo {author} {\bibfnamefont {F.}~\bibnamefont {Barocchi}}, \bibinfo {author} {\bibfnamefont {P.}~\bibnamefont {Mazzinghi}}, \bibinfo {author} {\bibfnamefont {V.}~\bibnamefont {Tognetti}},\ and\ \bibinfo {author} {\bibfnamefont {M.}~\bibnamefont {Zoppi}},\ }\bibfield  {title} {\bibinfo {title} {Determination of zone-boundary magnon energy and damping in $\mathrm{RbMnF}_{3}$ by means of light scattering experiments},\ }\href {https://doi.org/10.1016/0038-1098(78)90222-3} {\bibfield  {journal} {\bibinfo  {journal} {Solid State Commun.}\ }\textbf {\bibinfo {volume} {25}},\ \bibinfo {pages} {241} (\bibinfo {year} {1978})}\BibitemShut {NoStop}%
\bibitem [{\citenamefont {Pisarev}\ \emph {et~al.}(1973)\citenamefont {Pisarev}, \citenamefont {Moch},\ and\ \citenamefont {Dugautier}}]{pisarev1973light}%
  \BibitemOpen
  \bibfield  {author} {\bibinfo {author} {\bibfnamefont {R.~V.}\ \bibnamefont {Pisarev}}, \bibinfo {author} {\bibfnamefont {P.}~\bibnamefont {Moch}},\ and\ \bibinfo {author} {\bibfnamefont {C.}~\bibnamefont {Dugautier}},\ }\bibfield  {title} {\bibinfo {title} {{L}ight {S}cattering by {P}honons and {M}agnons in $\mathrm{NaNiF}_{3}$},\ }\href {https://doi.org/10.1103/PhysRevB.7.4185} {\bibfield  {journal} {\bibinfo  {journal} {Phys. Rev. B}\ }\textbf {\bibinfo {volume} {7}},\ \bibinfo {pages} {4185} (\bibinfo {year} {1973})}\BibitemShut {NoStop}%
\bibitem [{\citenamefont {Lockwood}\ \emph {et~al.}(1992)\citenamefont {Lockwood}, \citenamefont {Cottam},\ and\ \citenamefont {Baskey}}]{lockwood1992one}%
  \BibitemOpen
  \bibfield  {author} {\bibinfo {author} {\bibfnamefont {D.~J.}\ \bibnamefont {Lockwood}}, \bibinfo {author} {\bibfnamefont {M.~G.}\ \bibnamefont {Cottam}},\ and\ \bibinfo {author} {\bibfnamefont {J.~H.}\ \bibnamefont {Baskey}},\ }\bibfield  {title} {\bibinfo {title} {One-and two-magnon excitations in $\mathrm{NiO}$},\ }\href {https://doi.org/10.1016/0304-8853(92)90486-8} {\bibfield  {journal} {\bibinfo  {journal} {J. Magn. Magn. Mater.}\ }\textbf {\bibinfo {volume} {104}},\ \bibinfo {pages} {1053} (\bibinfo {year} {1992})}\BibitemShut {NoStop}%
\bibitem [{\citenamefont {Meloche}\ \emph {et~al.}(2007)\citenamefont {Meloche}, \citenamefont {Cottam},\ and\ \citenamefont {Lockwood}}]{meloche2007two}%
  \BibitemOpen
  \bibfield  {author} {\bibinfo {author} {\bibfnamefont {E.}~\bibnamefont {Meloche}}, \bibinfo {author} {\bibfnamefont {M.~G.}\ \bibnamefont {Cottam}},\ and\ \bibinfo {author} {\bibfnamefont {D.~J.}\ \bibnamefont {Lockwood}},\ }\bibfield  {title} {\bibinfo {title} {Two-magnon inelastic light scattering in the antiferromagnets $\mathrm{CoF}_{2}$ and $\mathrm{NiF}_{2}$: Experiment and theory},\ }\href {https://doi.org/10.1103/PhysRevB.76.104406} {\bibfield  {journal} {\bibinfo  {journal} {Phys. Rev. B}\ }\textbf {\bibinfo {volume} {76}},\ \bibinfo {pages} {104406} (\bibinfo {year} {2007})}\BibitemShut {NoStop}%
\bibitem [{\citenamefont {Bossini}\ \emph {et~al.}(2019)\citenamefont {Bossini}, \citenamefont {Dal~Conte}, \citenamefont {Cerullo}, \citenamefont {Gomonay}, \citenamefont {Pisarev}, \citenamefont {Borovsak}, \citenamefont {Mihailovic}, \citenamefont {Sinova}, \citenamefont {Mentink}, \citenamefont {Rasing},\ and\ \citenamefont {Kimel}}]{bossini2019laser}%
  \BibitemOpen
  \bibfield  {author} {\bibinfo {author} {\bibfnamefont {D.}~\bibnamefont {Bossini}}, \bibinfo {author} {\bibfnamefont {S.}~\bibnamefont {Dal~Conte}}, \bibinfo {author} {\bibfnamefont {G.}~\bibnamefont {Cerullo}}, \bibinfo {author} {\bibfnamefont {O.}~\bibnamefont {Gomonay}}, \bibinfo {author} {\bibfnamefont {R.~V.}\ \bibnamefont {Pisarev}}, \bibinfo {author} {\bibfnamefont {M.}~\bibnamefont {Borovsak}}, \bibinfo {author} {\bibfnamefont {D.}~\bibnamefont {Mihailovic}}, \bibinfo {author} {\bibfnamefont {J.}~\bibnamefont {Sinova}}, \bibinfo {author} {\bibfnamefont {J.~H.}\ \bibnamefont {Mentink}}, \bibinfo {author} {\bibfnamefont {T.}~\bibnamefont {Rasing}},\ and\ \bibinfo {author} {\bibfnamefont {A.~V.}\ \bibnamefont {Kimel}},\ }\bibfield  {title} {\bibinfo {title} {Laser-driven quantum magnonics and terahertz dynamics of the order parameter in antiferromagnets},\ }\href {https://doi.org/10.1103/PhysRevB.100.024428} {\bibfield  {journal} {\bibinfo  {journal} {Phys. Rev. B}\ }\textbf {\bibinfo {volume} {100}},\
  \bibinfo {pages} {024428} (\bibinfo {year} {2019})}\BibitemShut {NoStop}%
\bibitem [{sup()}]{supp_mat}%
  \BibitemOpen
  \href {https://journals.aps.org/supplemental/AAA/BBB} {}\bibinfo {howpublished} {\url{https://journals.aps.org/supplemental/AAA/BBB}}\BibitemShut {NoStop}%
\bibitem [{\citenamefont {Fedianin}\ \emph {et~al.}(2023)\citenamefont {Fedianin}, \citenamefont {Kalashnikova},\ and\ \citenamefont {Mentink}}]{fedianin2022selection}%
  \BibitemOpen
  \bibfield  {author} {\bibinfo {author} {\bibfnamefont {A.~E.}\ \bibnamefont {Fedianin}}, \bibinfo {author} {\bibfnamefont {A.~M.}\ \bibnamefont {Kalashnikova}},\ and\ \bibinfo {author} {\bibfnamefont {J.~H.}\ \bibnamefont {Mentink}},\ }\bibfield  {title} {\bibinfo {title} {Selection rules for ultrafast laser excitation and detection of spin correlation dynamics in a cubic antiferromagnet},\ }\href {https://doi.org/10.1103/PhysRevB.107.144430} {\bibfield  {journal} {\bibinfo  {journal} {Phys. Rev. B}\ }\textbf {\bibinfo {volume} {107}},\ \bibinfo {pages} {144430} (\bibinfo {year} {2023})}\BibitemShut {NoStop}%
\bibitem [{\citenamefont {Fabiani}\ and\ \citenamefont {Mentink}(2022)}]{fabiani2022parametrically}%
  \BibitemOpen
  \bibfield  {author} {\bibinfo {author} {\bibfnamefont {G.}~\bibnamefont {Fabiani}}\ and\ \bibinfo {author} {\bibfnamefont {J.~H.}\ \bibnamefont {Mentink}},\ }\bibfield  {title} {\bibinfo {title} {{P}arametrically driven {THz} magnon-pairs: {P}redictions toward ultimately fast and minimally dissipative switching},\ }\href {https://doi.org/10.1063/5.0080161} {\bibfield  {journal} {\bibinfo  {journal} {Appl. Phys. Lett.}\ }\textbf {\bibinfo {volume} {120}},\ \bibinfo {pages} {152402} (\bibinfo {year} {2022})}\BibitemShut {NoStop}%
\bibitem [{\citenamefont {Kimel}\ \emph {et~al.}(2009)\citenamefont {Kimel}, \citenamefont {Ivanov}, \citenamefont {Pisarev}, \citenamefont {Usachev}, \citenamefont {Kirilyuk},\ and\ \citenamefont {Rasing}}]{kimel2015inertia}%
  \BibitemOpen
  \bibfield  {author} {\bibinfo {author} {\bibfnamefont {A.~V.}\ \bibnamefont {Kimel}}, \bibinfo {author} {\bibfnamefont {B.~A.}\ \bibnamefont {Ivanov}}, \bibinfo {author} {\bibfnamefont {R.~V.}\ \bibnamefont {Pisarev}}, \bibinfo {author} {\bibfnamefont {P.~A.}\ \bibnamefont {Usachev}}, \bibinfo {author} {\bibfnamefont {A.}~\bibnamefont {Kirilyuk}},\ and\ \bibinfo {author} {\bibfnamefont {T.}~\bibnamefont {Rasing}},\ }\bibfield  {title} {\bibinfo {title} {Inertia-driven spin switching in antiferromagnets},\ }\href {https://doi.org/10.1038/nphys1369} {\bibfield  {journal} {\bibinfo  {journal} {Nat. Phys.}\ }\textbf {\bibinfo {volume} {5}},\ \bibinfo {pages} {727–731} (\bibinfo {year} {2009})}\BibitemShut {NoStop}%
\bibitem [{\citenamefont {Birss}(1964)}]{birss1964symmetry}%
  \BibitemOpen
  \bibfield  {author} {\bibinfo {author} {\bibfnamefont {R.~R.}\ \bibnamefont {Birss}},\ }\href@noop {} {\emph {\bibinfo {title} {Symmetry and magnetism}}}\ (\bibinfo  {publisher} {North-Holland Amsterdam},\ \bibinfo {year} {1964})\BibitemShut {NoStop}%
\bibitem [{\citenamefont {Mikhaylovskiy}\ \emph {et~al.}(2015)\citenamefont {Mikhaylovskiy}, \citenamefont {Hendry}, \citenamefont {Secchi}, \citenamefont {Mentink}, \citenamefont {Eckstein}, \citenamefont {Wu}, \citenamefont {Pisarev}, \citenamefont {Kruglyak}, \citenamefont {Katsnelson}, \citenamefont {Rasing},\ and\ \citenamefont {Kimel}}]{mikhaylovskiy2015ultrafast}%
  \BibitemOpen
  \bibfield  {author} {\bibinfo {author} {\bibfnamefont {R.~V.}\ \bibnamefont {Mikhaylovskiy}}, \bibinfo {author} {\bibfnamefont {E.}~\bibnamefont {Hendry}}, \bibinfo {author} {\bibfnamefont {A.}~\bibnamefont {Secchi}}, \bibinfo {author} {\bibfnamefont {J.~H.}\ \bibnamefont {Mentink}}, \bibinfo {author} {\bibfnamefont {M.}~\bibnamefont {Eckstein}}, \bibinfo {author} {\bibfnamefont {A.}~\bibnamefont {Wu}}, \bibinfo {author} {\bibfnamefont {R.~V.}\ \bibnamefont {Pisarev}}, \bibinfo {author} {\bibfnamefont {V.~V.}\ \bibnamefont {Kruglyak}}, \bibinfo {author} {\bibfnamefont {M.~I.}\ \bibnamefont {Katsnelson}}, \bibinfo {author} {\bibfnamefont {{\relax Th}.}~\bibnamefont {Rasing}},\ and\ \bibinfo {author} {\bibfnamefont {A.~V.}\ \bibnamefont {Kimel}},\ }\bibfield  {title} {\bibinfo {title} {Ultrafast optical modification of exchange interactions in iron oxides},\ }\href {https://doi.org/10.1038/ncomms9190} {\bibfield  {journal} {\bibinfo  {journal} {Nat. Commun.}\ }\textbf {\bibinfo {volume} {6}},\ \bibinfo
  {pages} {8190} (\bibinfo {year} {2015})}\BibitemShut {NoStop}%
\bibitem [{\citenamefont {Atxitia}\ \emph {et~al.}(2013)\citenamefont {Atxitia}, \citenamefont {Ostler}, \citenamefont {Barker}, \citenamefont {Evans}, \citenamefont {Chantrell},\ and\ \citenamefont {Chubykalo-Fesenko}}]{Atxitia-PRB2013}%
  \BibitemOpen
  \bibfield  {author} {\bibinfo {author} {\bibfnamefont {U.}~\bibnamefont {Atxitia}}, \bibinfo {author} {\bibfnamefont {T.}~\bibnamefont {Ostler}}, \bibinfo {author} {\bibfnamefont {J.}~\bibnamefont {Barker}}, \bibinfo {author} {\bibfnamefont {R.~F.~L.}\ \bibnamefont {Evans}}, \bibinfo {author} {\bibfnamefont {R.~W.}\ \bibnamefont {Chantrell}},\ and\ \bibinfo {author} {\bibfnamefont {O.}~\bibnamefont {Chubykalo-Fesenko}},\ }\bibfield  {title} {\bibinfo {title} {Ultrafast dynamical path for the switching of a ferrimagnet after femtosecond heating},\ }\href {https://doi.org/10.1103/PhysRevB.87.224417} {\bibfield  {journal} {\bibinfo  {journal} {Phys. Rev. B}\ }\textbf {\bibinfo {volume} {87}},\ \bibinfo {pages} {224417} (\bibinfo {year} {2013})}\BibitemShut {NoStop}%
\bibitem [{\citenamefont {Hellsvik}\ \emph {et~al.}(2016)\citenamefont {Hellsvik}, \citenamefont {Mentink},\ and\ \citenamefont {Lorenzana}}]{Hellsvik-PRB2016}%
  \BibitemOpen
  \bibfield  {author} {\bibinfo {author} {\bibfnamefont {J.}~\bibnamefont {Hellsvik}}, \bibinfo {author} {\bibfnamefont {J.~H.}\ \bibnamefont {Mentink}},\ and\ \bibinfo {author} {\bibfnamefont {J.}~\bibnamefont {Lorenzana}},\ }\bibfield  {title} {\bibinfo {title} {Ultrafast cooling and heating scenarios for the laser-induced phase transition in {CuO}},\ }\href {https://doi.org/10.1103/PhysRevB.94.144435} {\bibfield  {journal} {\bibinfo  {journal} {Phys. Rev. B}\ }\textbf {\bibinfo {volume} {94}},\ \bibinfo {pages} {144435} (\bibinfo {year} {2016})}\BibitemShut {NoStop}%
\bibitem [{\citenamefont {Hinzke}\ \emph {et~al.}(2015)\citenamefont {Hinzke}, \citenamefont {Atxitia}, \citenamefont {Carva}, \citenamefont {Nieves}, \citenamefont {Chubykalo-Fesenko}, \citenamefont {Oppeneer},\ and\ \citenamefont {Nowak}}]{Hinzke-PRB2015}%
  \BibitemOpen
  \bibfield  {author} {\bibinfo {author} {\bibfnamefont {D.}~\bibnamefont {Hinzke}}, \bibinfo {author} {\bibfnamefont {U.}~\bibnamefont {Atxitia}}, \bibinfo {author} {\bibfnamefont {K.}~\bibnamefont {Carva}}, \bibinfo {author} {\bibfnamefont {P.}~\bibnamefont {Nieves}}, \bibinfo {author} {\bibfnamefont {O.}~\bibnamefont {Chubykalo-Fesenko}}, \bibinfo {author} {\bibfnamefont {P.~M.}\ \bibnamefont {Oppeneer}},\ and\ \bibinfo {author} {\bibfnamefont {U.}~\bibnamefont {Nowak}},\ }\bibfield  {title} {\bibinfo {title} {Multiscale modeling of ultrafast element-specific magnetization dynamics of ferromagnetic alloys},\ }\href {https://doi.org/10.1103/PhysRevB.92.054412} {\bibfield  {journal} {\bibinfo  {journal} {Phys. Rev. B}\ }\textbf {\bibinfo {volume} {92}},\ \bibinfo {pages} {054412} (\bibinfo {year} {2015})}\BibitemShut {NoStop}%
\end{thebibliography}%

\end{document}